\documentclass[onecolumn]{els-mrw} 
\usepackage{amsmath,amssymb,amsfonts,amsthm,makeidx,graphicx}
\usepackage{txfonts}
\usepackage{helvet}
\usepackage{physics}
\usepackage{bbm}
\newcommand{\beq}{\begin{equation}}
\newcommand{\eeq}{\end{equation}}

\renewcommand{\d}{\mathrm{d}}

\def\oper{{\mathchoice{\rm 1\mskip-4mu l}{\rm 1\mskip-4mu l}
		{\rm 1\mskip-4.5mu l}{\rm 1\mskip-5mu l}}}
\newcommand{\cL}{{\cal L}}

\newcommand{\Psidag}{\Psi^{\ddagger}}
\begin{document}
\chapter{What We Talk About When We Talk About Dissipative Quantum Chaos}

\label{chap1}
\author[1]{Lucas S\'a}%
\author[2]{Pedro Ribeiro}%
\author[3]{Sergey Denisov}%
\address[1]{\orgname{University of Cambridge}, \orgdiv{TCM Group, Cavendish Laboratory, Ray Dolby Centre}, \orgaddress{JJ Thomson Avenue, Cambridge, CB3 0US UK}}
\address[2]{\orgname{Universidade de Lisboa}, \orgdiv{CeFEMA-LaPMET, Physics Department, Instituto Superior T\'ecnico}, \orgaddress{1049-001 Lisboa, Portugal}}
\address[3]{\orgname{OsloMet – Oslo Metropolitan University}, \orgdiv{Department of Computer Science}, \orgaddress{N-0130 Oslo, Norway}}

\articletag{Chapter Article tagline: April 16, 2026}

\maketitle

\begin{glossary}[Keywords]
Open quantum systems, quantum chaos, random matrix theory, quantum computing
\end{glossary}

\begin{abstract}[Abstract]
Dissipative quantum chaos is an emerging theory that is expected to extend the ideas, concepts, and methodology of conventional Hamiltonian quantum chaos from coherent evolution to open quantum dynamics. The new theory should provide a set of tools to distinguish
chaotic open quantum systems from integrable ones, as well as quantitative measures of their chaoticity (or, conversely, integrability).  
The foundations of this theory were laid in the late 1980s, and from the very start it was clear that, like its Hamiltonian predecessor, it had to be based on the spectral properties of the operators governing open quantum evolution. After these first steps, the field remained relatively quiet for many years and it is only over the last decade that the development of dissipative quantum chaos has received a strong boost, as confirmed by a large number of publications on this topic and, very recently, the first experiments performed to test its theoretical predictions. In this chapter, we review these recent developments and outline the basic foundations of dissipative quantum chaos.  
\end{abstract}

\section{Introduction}
 \label{intro}

\hskip 11.0cm {\sl
It seems to me we're just beginners.\footnote{Raymond Carver, {\sl What We Talk About When We Talk About Love} (1981).}
}
\medskip

Somewhere in the 1980s, there was a classical (by training) researcher with a strong background in nonlinear dynamics and dynamical chaos. After a decade of productive work on a remote island, the researcher decided to widen the horizon and learn quantum mechanics. During these studies, a question emerged that gradually took more and more of the researcher’s thoughts: Can the concepts of classical Hamiltonian chaos be related to the dynamics of quantum systems --- more complex than the harmonic oscillator or a freely propagating wave packet?

It is reasonable, thought the researcher, to begin with the key concept of dynamical chaos, that is sensitivity to initial conditions~\cite{EckmannRuelle1985,Lichtenberg1992}. This means that two trajectories in phase space, initially close to each other, diverge exponentially fast. This process is quantified with Lyapunov exponents~\cite{Ott2002}, the most important characteristics of classical chaos. But quantization of Lyapunov exponents is not straightforward, because the Schr\"odinger equation is linear. Yet one may use nonlinear functionals of the wave function, e.g., 
$Q$-representation~\cite{TodaIkeda1987} of the corresponding density operator or some metrics quantifying the distance between two states~\cite{HaakeWiedemannZyczkowski1992,ZyczkowskiSlomczynski1993}, and introduce a quantum version of Lyapunov exponents via them\footnote{See Chapter ``Quantum analogues of exponential sensitivity'' for more options.}. Another intuitive direction is \textit{reductio ad chaos}, that is, to use models that allow for a semiclassical limit, by, e.g.,  tuning down the value of the effective Planck constant~\cite{Casati1979} or increasing the maximal spin~\cite{HaakeKusScharf1987} or the number of interacting particles~\cite{EmaryBrandes2003}, and thus relate the original quantum system to a classical Hamiltonian one that can be classified as chaotic or integrable~\cite{Lichtenberg1992}. There is also an appealing visualization-based approach, which tracks the evolution of an initially localized wave packet by monitoring its Wigner or Husimi function~\cite{Berry1977,Balazs1984,Takahashi1986}.

Over the years, the researcher explored all these ideas. Although this exploration revealed many interesting, deep, and thought-provoking connections between classical chaos and quantum unitary evolution, the researcher remained somewhat unsatisfied because the approaches did not merge into a universal framework.

The real story of quantum chaos (QC)~\cite{Stockmann1999,haakeQuantumSignaturesChaos2019}, being the result of the collective efforts of many researchers, is much more satisfactory. Although it followed much the same path and explored the very same ideas as in the case of our lone researcher, it eventually arrived at a milestone, that is the realization that the connection between classical and quantum chaos is encoded in the spectral properties of Hamiltonians, the operators that generate coherent quantum evolution~\cite{Stockmann1999,Jensen1992}. This was the turning point, and the solid foundation on which a new theory could be built was provided by the, by then already well-developed, random-matrix theory (RMT)\footnote{See Chapter ``Quantum chaotic systems: a random-matrix approach''.}. A hallmark of the theory is the statistics of level spacings in the spectra of chaotic Hamiltonians, which, according to the Wigner surmise, show clear signatures of level repulsion absent in integrable quantum Hamiltonians. This was validated in early experiments with microwave billiards in the 1990s~\cite{Stockmann1999}. Later experiments on ultracold molecules~\cite{Zhou2010}, cold atoms~\cite{Khlebnikov2019}, and superconducting-qubit systems~\cite{Roushan2017} showed that by varying the system parameters, one can tune the spectral statistics between Poisson-like and Wigner-Dyson distributions.

We are not going to outline here the exciting history and main results of the QC theory, which is still flourishing and finds applications in many corners of modern quantum physics, ranging from quantum computing to many-body localization\footnote{See Chapter ``Many-body localization''.}. We would like to direct the interested reader to other Chapters of this volume, discussing different aspects of QC.

\textit{Dissipative} quantum chaos (DQC) is an extension of conventional QC to open quantum dynamics. This extension is driven, among other things, by a natural inspiration to connect the classical and quantum worlds outside of the Hamiltonian setting. It is so natural that the lone researcher, once done with Hamiltonian QC, will surely continue in this direction. Dissipative nonlinear classical systems exhibit a plethora of phenomena absent in the Hamiltonian limit, such as strange attractors, different bifurcation scenarios, and the coexistence of attractors with different basins of attraction~\cite{Ott2002}. It is indeed exciting to find out how all this extends to the quantum world.

Although the first attempts to quantize dissipative chaos can be traced back to the 1980s (see Section~\ref{history}), DQC has received a major boost in the last decade. This revival has been driven by three main factors. First, there has been substantial development in the theory of open many-body systems, on theoretical, computational, and experimental fronts~\cite{Landi2022,Sieberer2025}.
This development brought a variety of tunable models 
to play with, analytically, numerically, and, in some cases, experimentally. 
Second, there has been rapid progress in non-Hermitian random matrix theory (RMT)~\cite{KhoruzhenkoSommers2011}, including new results on universality classes~\cite{bernard2002,magnea2008}, their spectral statistics, free-probability methods~\cite{MingoSpeicher2017}, and related analytical and numerical tools. At the same time, a strong link has developed between non-Hermitian random matrix theory and quantum information theory\footnote{See Chapter ``Quantum Chaos and Quantum Information''.} (the latter being already closely tied to open quantum systems~\cite{breuerTheoryOpenQuantum2010}).
Finally, the last decade has been marked by rapid progress in noisy intermediate-scale quantum (NISQ) computing~\cite{preskill2018}, which makes it possible to experiment with real-life many-body quantum systems that call for being treated as \textit{open}, thus providing a natural testbed for exploring DQC in vivo~\cite{luitz2021PRR,woldSpectraNoisyParameterized2025,Wold2025ExperimentalDetection}.

The goal of our contribution is to map the current landscape of DQC. It is a very dynamic landscape that changes rapidly, with new results and ideas emerging almost on a monthly basis. However, some concepts already seem to be established. Most importantly, there is a common understanding that, similarly to conventional Hamiltonian QC, the classification of open quantum systems into ``integrable'' (or ``regular'') and ``chaotic'' should be related to the spectral properties of the operators that define the system evolution, Lindbladians in the case of time-continuous evolution, and completely positive trace-preserving (CPTP) maps in the case of time-discrete evolution~\cite{Wolf2012}. There is also renewed interest in microscopic characteristics of evolution that are specific to open quantum systems, such as photon-emission statistics~\cite{espositoNonequilibriumFluctuationsFluctuation2009,ferrariDissipativeQuantumChaos2025}, which provide a keyhole into the underlying dissipative dynamics~\cite{Yusipov2020PhotonWaiting,Pankratov2025}. 
At the same time, there are already some open questions and issues that deserve to be articulated now, so that they attract attention and, hopefully, will be resolved.

The chapter is organized as follows. In the next Section~\ref{scene}, we introduce the key objects of analysis in DQC. In Section~\ref{history}, we give a brief chronology of the development of the theory. 
Section~\ref{RMT} presents the random-matrix-theory perspective on DQC.  Section~\ref{spectral} is devoted to spectral statistics, primarily correlations between eigenvalues of Lindblad operators and CPTP maps, and how they can be related to regular/chaotic classification. In Section~\ref{classification}, we discuss symmetry classifications relevant for non-Hermitian generators of open dynamics. We then turn to other signatures of DQC. In Section~\ref{semi}, we discuss semiclassical
signatures of DQC, while in Section~\ref{Lyapunov} we look at waiting-time distributions as a potential experimental route to DQC. Finally, Section~\ref{grobe_revision} presents a revision of the first results on the spectral aspects of DQC, reported in 1988, and we conclude the chapter in Section~\ref{outlook}, by addressing some --- already standing and pressing --- issues and controversies.

\section{Setting the Scene}
 \label{scene}

\begin{figure}
    \centering
    \includegraphics[width=0.5\linewidth]    %{Figure0new.pdf}
    {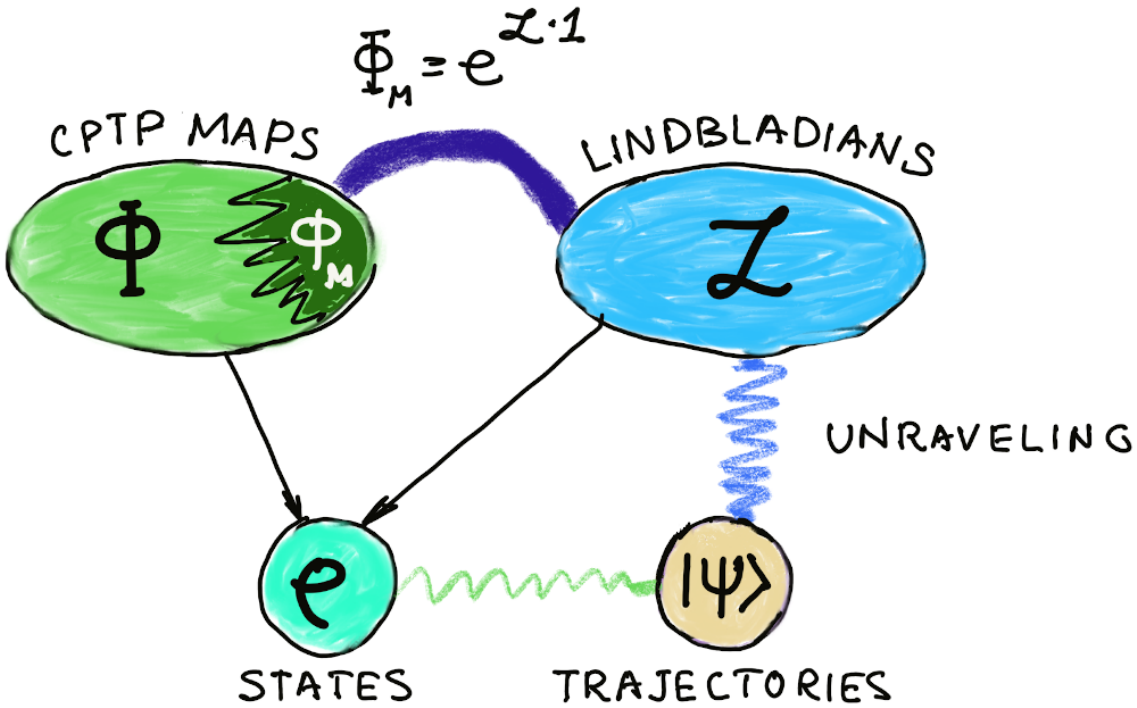}
    \caption{\textbf{Objects of dissipative quantum chaos}. Spectral properties of Lindblad operators $\cL$, CPTP maps $\Phi$, and quantum states $\rho(t)$ (especially asymptotic steady states $\rho_{\mathrm{ss}}$) are used to construct measures of integrability/chaoticity.
    Recently, ensembles of quantum trajectories generated by different unravelings~\cite{breuerTheoryOpenQuantum2010,wisemanQuantumMeasurementControl2010} of the same time-continuous open evolution have also started to be used for the same purpose.
    Note that all these objects are interrelated. For example, a density operator can be obtained by averaging over the ensemble of unravelled trajectories, and an important subclass of CPTP maps ~\cite{WolfEisertCubittCirac2008} is formed by maps $\Phi_M$ that can be represented as $\Phi_M=e^{\mathcal{L}\cdot 1}$.}
    \label{fig:objects}
\end{figure}

 The key objects of analysis in DQC are the operators that generate open evolution, Liouvillians $\cL$ (time-continuous evolution) and CPTP maps $\Phi$ (time-discrete evolution), as well as states $\rho(t)$ and trajectories $|\psi(t) \rangle$. All of these objects are interrelated; see Fig.~\ref{fig:objects}. We briefly describe them in the following.

Liouvillians are operators\footnote{In some texts they are called ``superoperators,'' to emphasize that their arguments are themselves quantum operators. We follow the standard mathematical terminology and regard them as operators acting on the vector space $M_N(\mathbb{C})$ of square complex matrices.} that generate continuous-time evolution of states,
\begin{equation}
    \dot{\rho}=\mathcal{L}\rho .
\end{equation}
Several forms of such operators have been considered in the literature~\cite{breuerTheoryOpenQuantum2010}. Although all of them are of interest in the context of DQC, here we focus  on Liouvillians of the celebrated Gorini--Kossakowski--Lindblad--Sudarshan (GKLS) form~\cite{goriniCompletelyPositiveDynamical1976,lindbladGeneratorsQuantumDynamical1976}.
\begin{equation}
\label{GKLS}
\cL = \cL_{\mathrm{H}}+\cL_{\mathrm{D}} = -\frac{i}{\hbar} [H,\rho] + \sum_{m,n=1}^{N^2}
K_{m,n}
\left(
F_{m}\rho F_{n}^{\dagger}
-\frac{1}{2}
\bigl\{
F_{m}^{\dagger}F_{n},\rho
\bigr\}
\right),
\end{equation}
where $N$ is the dimension of the Hilbert space, $\bigl\{\cdot,\cdot \bigr\}$ is the anticommutator 
and the operators $\{F_m\}_{m=1,...,N^2}$ form a Hilbert-Schmidt basis, $\mathrm{Tr}(F_nF_m^{\dagger})=\delta_{mn}$~\cite{Wolf2012}. Henceforth, we address such Liouvillians as `Lindblad operators' or simply `Lindbladians'.

The $N^2 \times N^2$ Kossakowski matrix $K$  of rank $R \leqslant N^2$ is Hermitian 
and positive semi-definite. Being such, it admits a representation $K=YY^{\dagger}$. Defining 
$L_{\alpha}= \sum_{m=1}^{N^2}
Y_{m\alpha}F_m,
~~
\alpha=1,\ldots,R$, we obtain the representation of a Lindbladian in terms of jump operators $L_{\alpha}$~\cite{breuerTheoryOpenQuantum2010,Wolf2012},
\begin{equation}
    \cL_{\mathrm{D}}(\rho) = \sum_{\alpha=1}^{R} \left(L_{\alpha}\rho L_{\alpha}^{\dagger} - \frac{1}{2} \{ L_{\alpha}^{\dagger}L_{\alpha},\rho \}\right).
    \label{eq:Lindblad_jump}
\end{equation}
The number of jump operators, $R$, is therefore set by the rank of the Kossakowski matrix.

Since $\cL$ is a Hermiticity-preserving operator, its spectrum is invariant under complex conjugation, $\ell\in\mathrm{spec}(\cL)\Rightarrow \ell^*\in\mathrm{spec}(\cL)$. Trace preservation implies that $\ell=0$ is an eigenvalue of $\cL$, and, in finite dimension $N$, the corresponding semigroup $\mathrm{exp}(t\cL)$ has at least one steady state $\rho_{ss}$, $\cL (\rho_{ss})=0$. 
All eigenvalues of a Lindblad generator belong to the closed left half-plane, $\mathrm{spec}(\cL)\subseteq\{\ell\in\mathbb{C}:\mathrm{Re}\,\ell\leq0\}$. Eigenvalues with $\mathrm{Re}\,\ell=0$ correspond to non-decaying modes, while eigenvalues with $\mathrm{Re}\,\ell<0$ correspond to decaying modes which do not contribute to the steady state.

A CPTP map $\Phi$ sets the time-discrete evolution $\rho_{n+1}=\Phi(\rho_n)$ and admits the Kraus representation \cite{breuerTheoryOpenQuantum2010,Wolf2012},
\begin{equation}
\Phi (\rho)= \sum_{\mu}^DK_{\mu}\rho K_{\mu}^{\dagger},
    \label{eq:CPTPmap}
\end{equation}
where $D$ is the Kraus rank defining the number of dissipative channels. This form enforces complete positivity, while normalization $\sum_{\mu}^DK_{\mu}^{\dagger}K_{\mu} = \oper$ ensures the preservation of the trace of the matrix on which the operator $\Phi$ is acting. 

Since $\Phi$ is Hermitian-preserving, its spectrum is invariant under complex conjugation, $\lambda\in\mathrm{spec}(\Phi)\Rightarrow \lambda^*\in\mathrm{spec}(\Phi)$. The quantum version of the Perron--Frobenius theorem~\cite{Wolf2012} implies that the spectral radius of any CPTP map is equal to one and that the spectrum is contained in the closed unit disk, $\mathrm{spec}(\Phi)\subseteq\{\lambda\in\mathbb{C}:|\lambda|\leq1\}$. Since $\Phi$ is trace-preserving, then, in finite dimension $N$, $1 \in\mathrm{spec}(\Phi)$ and there is at least one steady state $\rho_{ss}$, $\Phi (\rho_{ss})=\rho_{ss}$.

There are genetic ties connecting Lindbladians and maps. First, the dissipative part of a Lindbladian can be represented as 
\begin{equation}
\cL_{\mathrm{D}}(\rho) =  \Psi(\rho) -\frac{1}{2}\left( \Psidag(\oper)\rho + \rho  \Psidag(\oper)\right), 
  \label{eq:lindblad3}
\end{equation}
where $\Psi(\rho) = \sum_{\alpha=1}^{R} L_{\alpha}\rho L_{\alpha}^{\dagger} $ is a CP  (but not necessarily TP) map of rank $R$~\cite{Wolf2012} and $\Psi^\ddag$ denotes the map dual to it. Second, there is an important subclass of CPTP maps that can be obtained as exponentials of Lindbladians,  $\Phi_M=e^{1\cdot\mathcal{L}}$. They are also called ``Markovian'' or ``embeddable channels'' ~\cite{WolfEisertCubittCirac2008}.

Maps and Lindbladians are currently key objects of DQC and their spectral properties are used, similar to the case of Hamiltonians and unitary maps in Hamiltonian QC, as indicators of chaoticity/integrability; see, e.g., Refs.~\cite{grobeQuantumDistinctionRegular1988,akemannUniversalSignatureIntegrability2019,dahanClassicalQuantumChaos2022,pereiraDissipationInducedThresholdIntegrability2025,ferrariChaosThermalizationOpen2025}.

Quantum states themselves, represented by trace-one positive-semidefinite density operators $\rho(t)$, also serve as objects for DQC analysis~\cite{ProsenZnidaric2013,richter2025,villasenorBreakdownQuantumDistinction2024,rufoQuantumSemiClassicalSignatures2025,mondalTransientSteadyStateChaos2026}. The level spacing statistics of the steady state $\rho_{ss}$ has been proposed as a quantifier of chaoticity/integrability of the corresponding Lindbladian~\cite{ProsenZnidaric2013,richter2025}.

Finally, the evolution of the density operator can be unravelled~\cite{breuerTheoryOpenQuantum2010,wisemanQuantumMeasurementControl2010} into an ensemble of quantum trajectories, $\{|\psi_{\xi}(t) \rangle\} $, such that $\rho(t)=\mathbb{E}\!\left[\,\ket{\psi_\xi(t)}\!\bra{\psi_\xi(t)}\,\right]$.
Unrevealled trajectories were proposed as yet another means to construct new measures of the chaoticity~\cite{ferrariDissipativeQuantumChaos2025,Yusipov2020PhotonWaiting,yusipovQuantumLyapunovExponents2019,lumia2026}.

\section{Historical Overview}
\label{history}

Below, we provide a timeline of the development of DQC (which is far from exhaustive and highly subjective).

\medskip

\textbf{1981}. In his paper \emph{“Masers, lasers, and strange attractors”}~\cite{Oraevskii1981}, Oraevsky discussed the emergence of dissipative chaotic regimes “in quantum oscillators (masers and lasers)”. He considered the dynamics of the latter in terms of (semi)classical mean-field nonlinear dissipative 
equations for macroscopic amplitudes and populations. The treatment was therefore fully classical, yet a few lines in the concluding part can be used to justify citing this work as an early precursor of DQC. Namely, Oraevsky noted that the mean-field chaotic dynamics might be traced back to the spectra of the molecules and to the quantum interaction mechanisms that result in dissipative transitions between energy levels.

\textbf{1983}. In \emph{``Quantization of a Two-Dimensional Map with a Strange Attractor''}~\cite{Graham1983}, Graham constructed a quantized version of a dissipative two-dimensional map and obtained what today would be called a rank-two CPTP map~\cite{Wolf2012}. He derived the exact evolution of the density operator and explicitly evaluated the corresponding Wigner function. He found that, in the semiclassical limit, the steady-state Wigner function collapses onto the region of the phase space where the classical strange attractor is located. However, the latter, as a geometric set of non-integer Hausdorff dimension, does not survive in the quantum case, with its fractal structure being washed out by interference and uncertainty. Graham considered quantum versions of other classical dissipative maps in a series of subsequent works~\cite{Graham1984QuantumNoise,Graham1985}.

\textbf{1984}. In their paper~\cite{ElginSarkar1984}, Elgin and Sarkar considered a single-mode laser, modelled as a dilute gas of two-level atoms in a resonant cavity, using a quantum master equation of the GKLS form. They employed a moment-factorization approximation and, from the resulting equations for expectation values, claimed that, for weak quantum fluctuations, observables can show chaotic time dependence, while for stronger fluctuations their dynamics crosses over to periodic limit cycles or fixed points.

In his Comment~\cite{Graham1984}, Graham argued that this conclusion is incorrect and made an important observation: For a time-independent master equation, the density operator relaxes to a \textit{time-independent} invariant steady state $\rho_{ss}$, so, in the long-time limit, the observables must also become time independent and cannot keep evolving, neither periodically nor chaotically\footnote{Special care is needed, however, in non-generic cases, for example, when the zero eigenvalue is degenerate, when there is a decoherence-free (or dissipation-free) subspace, or, more generally, when the Liouvillian has eigenvalues $\mathrm{Re}\,\ell=0$ and the corresponding non-decaying modes may retain memory of the initial condition or even support persistent oscillations~\cite{Buca2019}.}. 
Therefore, the reported dynamics is an artifact of the approximation.

\textbf{1986}. Dittrich and Graham introduced a dissipative version~\cite{Dittrich1986} of quantum kicked rotor\footnote{For more information on this canonical model of QC see Chapter ``The quantum kicked rotor''.}~\cite{Casati1979}. Later works used the model and closely related kicked systems to explore dissipative quantum dynamics.

\textbf{1988}.  In their work \emph{``Quantum Distinction of Regular and Chaotic Dissipative Motion''}~\cite{grobeQuantumDistinctionRegular1988}, Grobe, Haake, and Sommers looked into spectral characteristics of the map defining time-discrete open quantum evolution and asked whether footprints of integrability and chaos can be detected, for a given set of parameters, in the spectral characteristics of the map. Their answer was affirmative. Since the spectrum of the map is complex, they generalized the standard notion of level spacing for real eigenvalues used in Hamiltonian QC to the complex case by defining it through the geometric distance between eigenvalues, $s_j=\min_{l\neq j}
\left|\lambda_j-\lambda_l\right|$. They also designed a two-dimensional version of the unfolding procedure and showed that, when the parameters correspond to the classically chaotic regime, the statistics of generalized level spacings follow that of the Ginibre ensemble (GinUE), whereas in the case of classically regular regime the statistics is close to the two-dimensional Poisson distribution, so that eigenvalues can be treated as an independent point process inside the unit disk.

This was a landmark in the development of DQC, for two reasons. First, for the first time the spectral approach was extended to open quantum evolution. Next, the latter was brought into contact with the theory of non-Hermitian random matrices. The work also contains seeds of other ideas that have recently reappeared in the DQC studies~\cite{akemannUniversalSignatureIntegrability2019,pereiraDissipationInducedThresholdIntegrability2025,navesLevelRepulsionFails2026}. 

It is worth giving a brief summary of this work here. Grobe, Haake, and Sommers  constructed a version of the dissipative quantum kicked rotor, represented by a $S$-spin particle, driven
with the time-periodic Hamiltonian
\begin{equation}
H(t)=H_0+H_1\sum_{n=0}^{\infty}\delta(t-n),~\mathrm{with}~~H_0=
pJ_z+\frac{k_0}{2S}J_z^2~~\mathrm{and}~~H_1
=
\frac{k_1}{2S}J_y^2.
\end{equation}
and subjected to dissipation described by the single jump operator $L=\frac{\Gamma}{2S}J_{-}$.
The object of analysis was the stroboscopic map,
\begin{equation}
\Phi(\rho) = e^{\cL_{H_0}+\cL_{D}}e^{\cL_{H_1}}\rho.
\label{eq:grobe}
\end{equation}
The dimension of the Hilbert space is $N=2S+1$ and the squared spin $J^2=S(S+1)$ is an invariant of the evolution. The map commutes with the operator $P(\rho)=R_z(\pi)\rho R_z^{\dagger}(\pi)$, $R_z(\pi)=e^{-i\pi J_z}$, so that the operator space, spanned by the basis $\{|m\rangle\langle n|\}$, splits into two invariant sectors, $m-n=\mathrm{even}$ and $m-n=\mathrm{odd}$, of dimensions $(S+1)^2+S^2$ and $2S(S+1)$, respectively (for a half-integer $S$ both are equal to $N^2/2$).

The model naturally allows one to approach the semiclassical limit by increasing $S$, so that the effective Planck constant scales as $\hbar_{\mathrm{eff}}\sim 1/S$; see Section~\ref{semi}. Grobe, Haake, and Sommers considered two regimes, a ``regular'' one, $k_1=0$, and a ``chaotic'' one, $k_1=8$. These labels refer to the corresponding dynamics in the semiclassical limit. As stated in the work, in the regular regime the motion on the classical spin sphere is dominated by simple attracting structures (e.g., limit cycles), whereas in the chaotic regime strong nonlinear kicking produces globally chaotic motion and, for nonzero damping, a strange attractor~\cite{Ott2002}.

\begin{figure}
    \centering
    \includegraphics[width=0.8\linewidth]
    {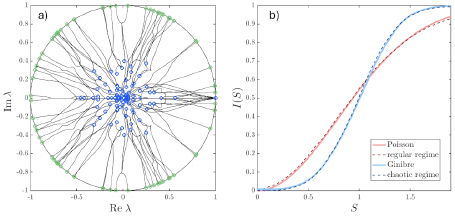}
\caption{\textbf{Reproduction of the results reported by Grobe, Haake, and Sommers in 1988~ \cite{grobeQuantumDistinctionRegular1988}}. (a) Motion of the complex eigenvalues of the stroboscopic map~\eqref{eq:grobe} as the damping strength $\Gamma$ is varied from $0$ (green circles) to $0.4$ (blue circles) in steps of $10^{-4}$. The eigenvalues are obtained for the even-parity sector in the chaotic regime. The parameters are $S=6$, $p=2$, $k_0=10$, $k_1=8$.
(b) Integrated nearest-neighbour spacing distribution $I(s)$. 
The spectra of the map are
computed for $S=10$, $\Gamma=0.1$, $p=2$, and $100$ equidistant values of
$k_0 \in [10,12]$. The regular case corresponds to $k_1=0$, while the
chaotic one corresponds to $k_1=8$. The distribution  for the regular regime is compared
with the Poisson distribution  for uncorrelated points in the plane, $I_{\mathrm{P}}(s)
=
1-\exp\!\left(-\pi s^2/4\right)$, whereas the distribution for the
chaotic regime is compared with the distribution obtained by sampling over complex $N^2 \times N^2$ Ginibre matrices. }
\label{fig:Figure1}
\end{figure}

Figure~\ref{fig:Figure1} presents our reproduction of the key results reported in Ref.~\cite{grobeQuantumDistinctionRegular1988}. The left panel shows the dynamics of the eigenvalues of the stroboscopic map under a gradual increase of the dissipation strength $\Gamma$, in the chaotic regime. The eigenvalue trajectories reveal nontrivial dynamics, including collisions and the consequent formation of a cluster around zero (see Section~\ref{random_maps} for a related discussion). These results served as inspiration for a recent work~\cite{pereiraDissipationInducedThresholdIntegrability2025}.
The right panel presents the main finding (as discussed above). 
These results are of foundational importance for DQC and therefore deserve an even more detailed discussion; see Section~\ref{grobe_revision}.

Next year, in 1989, Grobe and Haake coined the term \emph{dissipative quantum chaos} by publishing the work \emph{``Universality of cubic-level repulsion for dissipative quantum chaos''}~\cite{GrobeHaake1989}. They analyzed the level repulsion of complex eigenvalues in greater detail and also discussed, for the first time, symmetries of CPTP maps.

\textbf{1990 -- 2008}. There followed two relatively quiet decades. A few works by Haake, Grobe, Graham, and Dittrich, extending their previous results, were published and, in 2001, Braun published a book summarizing his and his collaborators’ results under the title \emph{``Dissipative Quantum Chaos and Decoherence''}~\cite{Braun2001}.

\textbf{2009}. Bruzda, Cappellini, Sommers, and {\.Z}yczkowski published the paper \emph{``Random Quantum Operations''}~\cite{bruzdaRandomQuantumOperations2009}. They addressed two questions: (i) How to define and sample a random CPTP map and (ii) what the spectrum of such a map is. In retrospect, both questions turned out to be fundamental for DQC. The paper introduced several constructions of random CPTP maps of a given rank, based on the Choi-Jamio{\l}kowski isomorphism, the Kraus representation, and the Stinespring form~\cite{Wolf2012}. The main result of the work is spectral universality: In the asymptotic limit, apart from the leading eigenvalue $\lambda_1=1$, the eigenvalues concentrate inside a disk of radius $1/\sqrt{D}$, where $D$ is the Kraus rank of the map. In this sense, after dropping $\lambda_1=1$, the spectral density of a random quantum map is statistically close to that of a real Ginibre matrix of size $N^2-1$ with independent entries of zero mean and variance $1/[D(N^2-1)]$.

\textbf{2013}. Prosen and \v{Z}nidari\v{c}, in \emph{``Eigenvalue Statistics as an Indicator of Integrability of Nonequilibrium Density Operators''}~\cite{ProsenZnidaric2013}, analyzed the eigenvalue statistics of the nonequilibrium steady-state density operator, searching for a criterion to distinguish integrable and non-integrable regimes in open quantum systems. Their main conjecture was that Poissonian level statistics signals integrability, whereas in generic non-integrable cases the statistics follows the Wigner-Dyson distribution. This can be seen as a precursor to \emph{steady-state DQC}, which has attracted growing attention recently~\cite{ferrariDissipativeQuantumChaos2025,richter2025,mondalTransientSteadyStateChaos2026}.

\textbf{2019}. Denisov, Laptyeva, Tarnowski, Chru\'sci\'nski, and {\.Z}yczkowski published \emph{``Universal Spectra of Random Lindblad Operators''}, which addressed questions similar to those raised in the 2009 paper~\cite{bruzdaRandomQuantumOperations2009}, but now for random Lindblad operators\footnote{The problem of defining random Lindbladians and describing their spectral density was first addressed by Timm and Lange in 2011. Their results were presented in Lange's MSc thesis, written in German, which became available online only in 2020. Some of these results were later published in Ref.~\cite{langeRandomMatrixTheory2021}.}. It was followed by three works of other teams~\cite{can2019JPhysA,CanOganesyan2019,SaRibeiroProsen2020} (Ref.~\cite{SaRibeiroProsen2020} was posted on the arXiv the same year but published in 2020).

In the case of Lindblad operators, the situation is more complicated than in the case of maps, since the former consist of two parts, a Hamiltonian one and a dissipative one, and, before discussing spectral universalities, one has first to specify their relative weights. The Hamiltonian part can be sampled in a rather direct way by drawing the Hamiltonian from a Gaussian distribution, whereas the sampling of the dissipative part is much less unambiguous. One convenient route is to sample the positive semidefinite Kossakowski matrix. In the purely dissipative case, the properly rescaled spectral density of a random Lindbladian acquires a universal lemon-like shape; see Fig.~\ref{fig:Figure22}. We discuss spectra of random Lindbladians in Section~\ref{RL}.

The same year, Akemann, Kieburg, Mielke, and Prosen in \emph{``Universal Signature from Integrability to Chaos in Dissipative Open Quantum Systems''}~\cite{akemannUniversalSignatureIntegrability2019} and S\'a, Ribeiro, and Prosen in \emph{``Complex Spacing Ratios: A Signature of Dissipative Quantum Chaos''}~\cite{saComplexSpacingRatios2020a} (posted on the arXiv in 2019, but published in 2020) revived and refined the spectral approach proposed by Grobe, Haake, and Sommers~\cite{grobeQuantumDistinctionRegular1988}. While the latter analyzed discrete-time dissipative quantum maps, which are one-period propagators of the quantum kicked top model, and identified cubic level repulsion as the spectra of these operators, these works moved to Lindblad operators and considered a many-body open quantum system as an illustration. Ref.~\cite{saComplexSpacingRatios2020a} also introduced complex spacing ratios (CSRs), which have since become one of the most popular measures of integrability/chaos in open quantum systems; we discuss CSRs in Section~\ref{sec:local_stat}. 

All these 2019 works marked the beginning of a decade of revitalization of DQC, which continues to this day.

\section{Random Matrix Theory Perspective on Dissipative Quantum Chaos}
\label{RMT}

In this section, we discuss how to define the notion of “random operators” when it applies to the main objects of DQC, Lindbladians and CPTP maps. We also show that RMT, and especially the powerful apparatus of free-probability theory~\cite{MingoSpeicher2017}, allows one to evaluate the spectral densities of these random operators.

\subsection{Random Lindbladians}
\label{RL}

In line with the work of Bruzda, Cappellini, Sommers, and {\.Z}yczkowski~\cite{bruzdaRandomQuantumOperations2009} on random maps, this section addresses two questions: (i) how to construct a random Lindblad operator, and (ii) what spectral properties of such an operator are.

\subsubsection{Constructing random Lindbladians}

A construction of random $\cL$  naturally splits into two parts [see Eq.~\eqref{GKLS}], the construction of the Hamiltonian part
$\cL_H$ and that of the dissipative part $\cL_D$.

The definition of the random Hamiltonian part $\cL_H=-i[H,\cdot]$ is
straightforward. One simply samples $H$ from the Gaussian Orthogonal
Ensemble (GOE) or the Gaussian Unitary Ensemble (GUE)~\cite{mehta2004}. The spectrum of$\cL_H$ follows directly from the spectrum of $H$ (see the next subsection).

There are (at least) two intuitive ways to construct a random dissipative part, $\cL_D$, depending on the particular representation of this operator. We define the first construction using the representation given by Eq.~\eqref{GKLS}. The positive semidefinite Kossakowski matrix can be sampled as an $(N^2-1)\times (N^2-1)$ Wishart matrix,
$K=GG^{\dagger}$, where $G$ is drawn from the complex Ginibre ensemble (GinUE). There is a certain freedom in choosing the Hilbert-Schmidt basis. In Ref.~\cite{denisovUniversalSpectraRandom2019}, a traceless orthonormal
basis of generators of SU$(N)$ was used, with $F_1=\oper/\sqrt{N}$ and all remaining 
operators being traceless. Only the latter are used to construct the dissipative part.  This removes the identity component from the jump operators (which can otherwise result in a spurious appearance of the Hamiltonian part) and the Kossakowski matrix has dimension $N^2-1$. 
In order to have meaningful sampling, all realizations of $\cL_D$ should have the
same overall strength, defined by the trace of the Kossakowski matrix. Normalization $\mathrm{Tr}\,K=N$ was proposed~\cite{denisovUniversalSpectraRandom2019}, so that
$K=N\cdot GG^{\dagger}/\mathrm{Tr}(GG^{\dagger})$. 
Alternatively, one may use the matrix-unit basis
$\{E_{ab}=|a\rangle\langle b|\}$, $a,b=1,\ldots,N$. In this case,  a generic random jump operator sampled in this basis has a non-zero trace component. However, the latter scales down to $1/N^2$ so it becomes negligible in the thermodynamic limit. The Kossakowski matrix in this case has dimension $N^2$ and is sampled in the same way as before.  The rank $R$ of $K$, and therefore the number of jump operators, can be controlled directly. If the operator basis has dimension $d$, with
$d=N^2-1$ for the traceless SU$(N)$ generator basis and $d=N^2$ for the full
matrix-unit basis, one should sample a rectangular Ginibre matrix
$G\in\mathbb{C}^{d\times R}$.

The second construction of $\cL_D$ is based on the representation~\eqref{eq:lindblad3} and thus reduces to sampling a random CP map $\Psi$. The most natural way to do this is to use the Choi--Jamio{\l}kowski isomorphism~\cite{Wolf2012} and sample a random Choi operator $C_{\Psi}$. Similarly to the case of the Kossakowski matrix, $C$ can be sampled as $C_{\Psi}=GG^{\dagger}$. The rank $R$ can be controlled here in the same way, since the rank of the Choi operator is equal to the rank of the corresponding Kossakowski matrix. Also, the normalization reduces in this case to $\mathrm{Tr} C_{\Psi}=N$ so that $C_{\Psi}= N\cdot GG^\dagger/\mathrm{Tr}GG^{\dagger}.$
Other sampling methods, based on random sets of Kraus operators or on the Stinespring dilation, were proposed in Ref.~\cite{bruzdaRandomQuantumOperations2009}. 
Note that the sampling of a CP map is simpler than the sampling of a CPTP map~\cite{bruzdaRandomQuantumOperations2009} since no trace-preserving condition is imposed and, therefore, there is no need for the partial-trace normalization. It is noteworthy that if the CP map is trace preserving, then the dissipative Lindbladian reduces to the shifted map $\cL_D=\Phi-\oper_{N^2}$. The spectra of such operators are then trivially related to the spectra of the corresponding maps $\Phi$; see  Section~\ref{random_maps}.

We now construct a random Lindblad operator by combining two parts, $\cL_H$ and $\cL_D$. The only point to address is how to weight their relative contributions. We assume that the normalization of the dissipative part, defined above, is fixed, so it remains only to specify the normalization of the Hamiltonian part. Following Ref.~\cite{denisovUniversalSpectraRandom2019}, the Hamiltonian is normalized as $\mathrm{Tr}\,H^2=1/N$, and its relative
weight is defined by a positive constant $\alpha$,
\begin{equation}
\cL_{\alpha}=\alpha\cL_H+\cL_D.
\label{eq:fullLind}
\end{equation}
The general scaling
$\cL_{\alpha}\to\mu\cL_{\alpha}$, $\mu>0$, is trivial and fixes only the 
overall speed of the evolution.

\subsubsection{Spectra of random Lindbladians: RMT approach}

The RMT approach to the random Lindbladian given by Eq.~\eqref{eq:fullLind}  splits into three steps. First, we evaluate the spectrum of the random Hamiltonian part, $\cL_H$. Next, we handle the dissipative part, $\cL_D$. Finally, we address the general case, Eq.~\eqref{eq:fullLind}.

We start with the Hamiltonian part. If $H|\alpha\rangle=E_\alpha|\alpha\rangle$, then the operator basis
$|\alpha\rangle\langle\beta|$ diagonalizes 
$\cL_H=-i[H,\cdot]$. Indeed,
$\cL_H|\alpha\rangle\langle\beta|=-i(E_\alpha-E_\beta)
|\alpha\rangle\langle\beta|$, so the eigenvalues of $\cL_H$ are
$\ell_{\alpha\beta}=-i(E_\alpha-E_\beta)$. Hence, the spectrum of
$\cL_H$ lies on the imaginary axis and consists of all pairwise energy differences of the GUE spectrum, multiplied by $-i$. The $N$ diagonal operators $|\alpha\rangle\langle\alpha|$ give an $N$-fold degenerate zero eigenvalue. In the large-$N$ limit, the spectral density is therefore determined by the standard classical convolution $\rho_{\Delta}(\omega)=\int dE\,\rho_{\mathrm{sc}}(E)\rho_{\mathrm{sc}}(E-\omega)$, $\omega=E_\alpha-E_\beta$, of two Wigner semicircles,
$\rho_{\mathrm{sc}}(E)=\frac{1}{2\pi}\sqrt{4-E^2}$, $|E|\leq 2$. 
This convolution can be evaluated analytically,
\begin{equation}
\label{eq:delta_density_semicircle}
\rho_{\Delta}(\omega)
=
\frac{(16+\omega^2)\mathrm{E}\!\left(1-\frac{\omega^2}{16}\right)
-2\omega^2\mathrm{K}\!\left(1-\frac{\omega^2}{16}\right)}{6\pi^2},
\quad |\omega|\leq 4,
\qquad
\rho_{\Delta}(\omega)=0,
\quad |\omega|>4 .
\end{equation}
Here, $\mathrm{K}(x)$ and $\mathrm{E}(x)$ are complete elliptic integrals of
the first and second kind~\cite{GradshteynRyzhik}. Since $\ell=-i\omega$, the spectral density of
$\cL_H$ is supported on the imaginary axis. Writing $z=x+iy$, we obtain
\begin{equation}
\label{eq:LH_density_from_delta_density}
\rho_{\cL_H}(x,y)=\delta(x)\rho_{\Delta}(y),
\end{equation}
where the symmetry $\rho_{\Delta}(y)=\rho_{\Delta}(-y)$ has been used.

The RMT model of the dissipative part, Eq.~\eqref{eq:lindblad3}, can be constructed term by term. We rewrite it as
\begin{equation}
\cL_D(\rho) = \Psi(\rho) - \rho - \frac{1}{2} (X \rho + \rho X),
\label{eq:Lindblad_shifted}
\end{equation}
where $X$ is a Hermitian matrix. From the spectral RMT perspective, the map $\Psi$ can be represented by a real $N^2 \times N^2$ Ginibre matrix, i.e., a member of GinOE, up to a real non-negative outlier responsible for the stationary eigenvalue $\ell=0$, which is omitted in the bulk analysis. The second term on the rhs only shifts the spectrum by $-1$.

Writing $\Psi^\ddag(\oper)=\oper+X$, we obtain the centered Hermitian matrix $X=\Psi^\ddag(\oper)-\oper$. If $\Psi(\rho)=\sum_m\gamma_m V_m\rho V_m^\dagger$, then $\Psi^\ddag(\oper)=\sum_m\gamma_m V_m^\dagger V_m$. Since each contribution $V_m^\dagger V_m$ is positive and Wishart-like, then, after normalization, the mean value of the sum is $\oper$. Therefore, $X=\Psi^\ddag(\oper)-\oper$ describes the centered fluctuations around this mean. In the large-$N$ limit, these fluctuations are expected to be of GOE type. Finally, after applying the affine transform $\cL'_D = N(\cL_D+\oper_{N^2})$, we arrive at the following RMT model, up to the shift by $\oper_{N^2}$:
\begin{equation}
\label{eq:lemon_rmt_model}
{\cL}^{RMT}_{D} = G_R-
\left(
C\otimes \oper
+
\oper\otimes C
\right),
\end{equation}
where $G_R$ is drawn from GinOE and $C$ from GOE. The matrices are normalized as $\mathrm{Tr}\,G_R G_R^\dagger=N^2$, so that the spectrum of $G_R$ uniformly fills a disk of radius $1$, while $\mathrm{Tr}\,C^2=N/4$ ensures that the spectral density of $C$ is the Wigner semicircle with radius $1$.

The term $C\otimes \oper$ has a spectral density given by an $N$-fold degenerate Wigner semicircle, and the full second term is described by the classical self-convolution of this semicircle. This convolution was already evaluated above for $\cL_H$. The spectral support of ${\cL}^{RMT}_{D}$ is therefore obtained as the non-Hermitian free convolution~\cite{MingoSpeicher2017} of the disk generated by $G_R$ and the classical semicircle self-convolution generated by $C\otimes \oper+\oper\otimes C$.

\begin{figure}[t]
    \centering
    \includegraphics[width=0.95\linewidth]
    {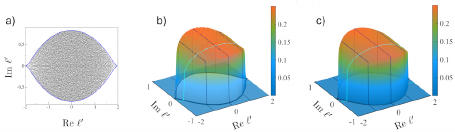}
      \caption{
\textbf{Spectral density of random Lindblad operators in the purely dissipative limit}. (a) Rescaled eigenvalues $\ell'=N(\ell+1)$ of a single random Lindblad operator for $N=50$ (the stationary eigenvalue $\ell_1=0$ is omitted). The solid contour shows the asymptotic boundary of the universal lemon-shaped support, Eqs.~\eqref{eq:lemon_boundary_condition} and~\eqref{eq:lemon_G_function}. 
(b,c) Probability density $P(\operatorname{Re}\ell',\operatorname{Im}\ell')$ of the rescaled eigenvalues. Panel (b) presents the density obtained by sampling $10^4$ random Lindblad operators for $N=100$, while panel (c) shows the results obtained with the random-matrix model, Eq.~\eqref{eq:lemon_rmt_model}. The dark blue contours indicate the lemon-shaped support. Note that the real eigenvalues were discarded when sampling random Lindbladians, since they form an additional narrow crest along the real axis.
}
    \label{fig:Figure22}
\end{figure}

The spectral support of Eq.~\eqref{eq:lemon_rmt_model} was evaluated analytically using the quaternionic extension of free probability~\cite{denisovUniversalSpectraRandom2019}.
It was shown that the boundary of the spectrum is determined by the solution of
\begin{equation}
\label{eq:lemon_boundary_condition}
\operatorname{Im}\!\left[z+G(z)\right]=0,
\end{equation}
where $z$ is a complex variable and
\begin{equation}
\label{eq:lemon_G_function}
G(z)
=
2z
-
\frac{2z}{3\pi}
\left[
(4+z^2)\mathrm{E}\!\left(\frac{4}{z^2}\right)
+
(4-z^2)\mathrm{K}\!\left(\frac{4}{z^2}\right)
\right].
\end{equation}
This boundary has a characteristic lemon-like shape and is in excellent agreement with the results of numerical sampling of random $\cL_D$; see Fig.~\ref{fig:Figure22}(a). Moreover, the agreement persists at the more detailed level of the spectral density; see Fig.~\ref{fig:Figure22}(b,c). The spectral density inside the lemon was analytically evaluated in Ref.~\cite{langeRandomMatrixTheory2021}.

Now we turn to the general case, Eq.~\eqref{eq:fullLind}.
From the RMT perspective, the correction produced by the Hamiltonian part, $\cL_H=-i[H,\cdot]$, can be absorbed into the second term in the rhs of Eq.~\eqref{eq:lemon_rmt_model}. With the standard column-vectorization convention,
$\mathrm{vec}(A\rho B)=(B^{\mathsf T}\otimes A)\mathrm{vec}(\rho)$, the Hamiltonian part is represented as $
-i[H,\rho]
 \rightarrow 
-i(\oper\otimes H-H^{\mathsf T}\otimes \oper)$.
Therefore, after combining this term with the dissipative contribution, the deterministic part of the RMT model becomes
$
-
\left(
W^*\otimes \oper
+
\oper\otimes W
\right),
~
W=C+i\alpha H 
$,
with $C$ being a member of GOE, while $H$ is of  GUE.

\begin{figure}[t]
    \centering
    \includegraphics[width=0.99\linewidth]%{figures/Figure3new.pdf}
    {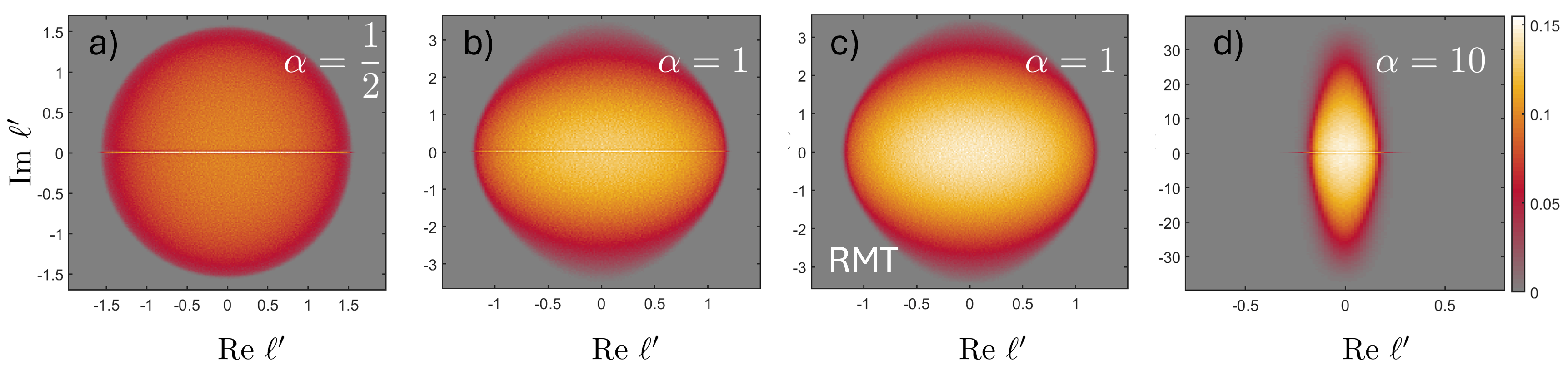}
\caption{Spectral density of random Lindblad operators, Eq.~\eqref{eq:fullLind}, for different values of the weight $\alpha$ of the Hamiltonian contribution~\cite{denisovUniversalSpectraRandom2019}. The panels show the density of the rescaled eigenvalues $\ell'=N(\ell+1)$. Panel (a) corresponds to the case $\alpha=1/2$, where the spectral support is expected to become a disk. Panel (b) shows the result for $\alpha=1$, which should be compared with the result for the RMT model shown in panel (c). %Adapted from Ref.~\cite{denisovUniversalSpectraRandom2019}.
}
    \label{fig:Figure222}
\end{figure}

The spectral densities sampled with the RMT model for different values of $\alpha$ are presented in Fig.~\ref{fig:Figure222}. By increasing $\alpha$, we first round and then squeeze the dissipative lemon. The value $\alpha=1/2$ is special in the present normalization. Indeed, the GOE matrix $C$ is normalized as $\mathrm{Tr}C^2=N/4$, whereas the Hamiltonian matrix $H$ is normalized as $\mathrm{Tr}H^2=N$. Hence the Hermitian part $C$ and the anti-Hermitian part $i\alpha H$ have equal effective variance precisely when $
\alpha^2\mathrm{Tr}H^2=\mathrm{Tr}C^2,
$
which gives $\alpha=1/2$. At this point, $W=C+i\alpha H$ belongs to the circular Ginibre-type universality class, and its spectral support is a disk~\cite{KhoruzhenkoSommers2011}. Consequently, the tensor contribution $W^*\otimes \oper+\oper\otimes W$ is governed, at the level of the support, by the classical convolution of two disks. This support is again rotationally symmetric. Together with the disk generated by the GinOE contribution $G_R$, this symmetry implies that the final spectral support is circular. This is in excellent agreement with the numerical sampling; see Fig.~\ref{fig:Figure222}(a). The bright line along the real axis corresponds to the concentration of real eigenvalues, which is typical of the GinOE ensemble~\cite{Tarnowski2022}.

Away from special points $\alpha=0$ and $\alpha=1/2$, an analytic evaluation of the spectral boundary and density is substantially more challenging~\cite{hontarenko}. 

\subsubsection{Random Lindbladians with non-maximal rank}

The Wishart construction allows one to control the rank of random Kossakowski matrices and thus the number of jump operators. Let $d$ be the dimension of the operator basis, with $d=N^2-1$ for the traceless SU$(N)$ basis and $d=N^2$ for the full matrix-unit basis. A Kossakowski matrix of rank $R$ can be sampled by taking a rectangular Ginibre matrix $G\in\mathbb{C}^{d\times R}$ and setting
\begin{equation}
K=\frac{N\,GG^\dagger}{\mathrm{Tr}(GG^\dagger)} .
\end{equation}
Then $\mathrm{rank}\,K = R$, and after diagonalizing $K$ the dissipator contains at most $R$ independent jump operators. 

Reducing $R$ while keeping $\mathrm{Tr}\,K=N$ fixed changes the scaling of the spectral support. In the full-rank case, $R\simeq d\simeq N^2$, the nontrivial spectrum of the purely dissipative Lindbladian is described by the shifted and rescaled variable
$
\ell'=N(\ell+1)
$,
which gives the universal lemon-shaped support~\cite{denisovUniversalSpectraRandom2019}. More generally, the scale of the random CP part and of the fluctuations of the loss term is controlled by the effective number of channels. At the heuristic level, the natural scaling is therefore
$\ell'=\sqrt{R}\,(\ell+1)$. 
If $R=r N^2$, where the realtive rank $0<r<1$, the support plotted in the variable $N(\ell+1)$ is expanded by a factor of the order $r^{-1/2}$.

If the rank $R$ of the Kossakowski matrix remains finite as $N\to\infty$, the lemon shape of the spectral support is no longer expected in the thermodynamic limit. The loss operator no longer self-averages to a scalar contribution and the problem crosses over to the few-channel regime, which has been studied in several works. For example, Can showed that a purely dissipative random Lindbladian with a single jump operator has a spectral gap closing in the thermodynamic limit, whereas adding more jump operators or adding a Hamiltonian part can open a gap that remains finite as $N\to\infty$~\cite{can2019JPhysA}. Can, Oganesyan, Orgad, and Gopalakrishnan also studied random Hamiltonians with random jump operators and found a gapped large-$N$ spectrum, together with a sharp transition between a weak-dissipation regime, where the nonzero eigenvalues form a continuum, and a strong-dissipation regime, where the asymptotic decay rate becomes an isolated mid-gap mode~\cite{CanOganesyan2019}. S{\'a}, Ribeiro, and Prosen considered random Lindbladians with a random Hamiltonian and a finite number of independent jump operators. They found that the global spectral features, the gap, and the steady-state density matrix obey distinct scaling regimes as functions of dissipation strength and system size. In particular, for two or more channels, the gap grows with system size, while the steady-state spectrum crosses over from Poissonian statistics at weak dissipation to Wigner-Dyson statistics at strong dissipation~\cite{SaRibeiroProsen2020}.

The genuinely
sparse Lindblad limit, in which the number of jump operators is of order
$O(1)$, i.e., it is fixed and does not scale with the Hilbert-space dimension
$N$, remains mainly analytically unexplored.
In this limit the dissipators cover only a very small
fraction of structured directions in operator space, and therefore spectral statistics of the corresponding Lindblad operator 
cannot be expected to follow directly from full-rank arguments. Establishing spectral universality in this limit is an
open and challenging problem.

\subsection{Random Maps}
\label{random_maps}
We treat CPTP maps as non-Hermitian operators acting on the space of complex matrices. The construction of random CPTP maps and the analysis of their spectra were developed in Ref.~\citep{bruzdaRandomQuantumOperations2009}. We use these results as a starting point.

\subsubsection{Random Kraus maps as minimal models of generic open
dynamics}\label{random-kraus-maps-as-minimal-models-of-generic-open-dynamics}

A natural construction of a random CPTP map starts, e.g., from sampling a Haar-random unitary acting on the system and an $M$-dimensional environment, followed by tracing out the environment. Equivalently, one may sample a random Choi matrix subject to the partial-trace constraint~\citep{bruzdaRandomQuantumOperations2009} and then obtain the corresponding random map via the Choi-Jamio{\l}kowski isomorphism~\cite{Wolf2012}. Both constructions induce a natural measure on the convex set of CPTP maps. From the DQC perspective, such random map ensembles play a role analogous to Wigner--Dyson ensembles in Hamiltonian chaos, i.e., they provide \textit{typical} operators for which one may expect random-matrix universality, both in spectral statistics of the operator and, in suitable regimes, in the corresponding steady-state density operator.

\begin{figure}[t]
    \centering
    \includegraphics[width=0.85\linewidth]{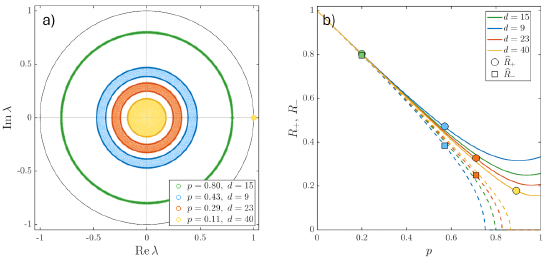}
    \caption{\textbf{
Spectra of diluted unitaries~\citep{saSpectralTransitionsUniversal2020}}. In this type of random CPTP maps, Eq.~\eqref{eq:diluted_unitaries}, $p$ controls the strength of the dissipative component and $d$ is the number of decay channels. (a) Spectra of single realizations of $\mathcal{E}_p$ for different values of $p$ and $d$, $N=50$. At weak dissipation the spectrum forms a ring, while increasing $p$ reduces the inner radius and eventually results in a closure to a disk. (b) Inner and outer radii, $R_-(p)$ and $R_+(p)$, of the spectral support. Symbols show the estimates obtained from random Kraus maps sampled for the corresponding values of $p$ and $d$. For each map realization, the outer and inner radii were estimated as the averages over the moduli of the $10$ outermost and $10$ innermost nontrivial eigenvalues, respectively. The curves show the large-$N$ random-matrix prediction, Eq.~\eqref{eq:radii_diluted_unitaries}. 
}
    \label{fig:Figure5}
\end{figure}

Similar to the case of Lindbladians discussed in the previous section, it is essential that random maps can be realized not only
as fully nonlocal operators acting on the whole operator space at
once, but also as \emph{local} random Kraus circuits built from few‑site
CPTP blocks arranged in a brickwork pattern\footnote{It is noteworthy that local Lindbladians do not generally generate strictly local maps. Exponentiation smears out the locality of the support and the locality survives only in the Lieb--Robinson sense~\cite{poulin2010}.}
\citep{saSpectralTransitionsUniversal2020}. In the ``0D'' version, the
map acts as if all degrees of freedom
were fully connected, while in the ``1D'' version, the corresponding Hilbert space is the space of a chain of qudits and each time step is generated by layers of two‑site
Kraus gates with the same local structure repeated across space and time. A key result of
Ref.~\citep{saSpectralTransitionsUniversal2020} is that the local
circuits display spectral supports, spectral gaps, and steady‑state
statistics that are qualitatively similar to those of their
nonlocal counterparts. This observation justifies treating the simpler
nonlocal random Kraus maps as efficient minimal models for generic
interacting open dynamics in DQC, while retaining a clear interpretation
in terms of non‑unitary quantum circuits.

\subsubsection{Diluted unitary circuits and the ring-disk
transition}\label{diluted-unitary-circuits-and-the-annulusdisk-transition}

In Ref.~\citep{saSpectralTransitionsUniversal2020} a family of \emph{diluted unitary} maps, in which the Kraus operators are designed so that the overall map can be written as a mixture of a unitary map and a strongly mixing random component, was introduced and analysed. In the fully connected, or ``0D'', setting one has
\begin{equation}
\label{eq:diluted_unitaries}
\mathcal{E}_p=(1-p)\mathcal{U}+p\Phi,
\end{equation}
where $\mathcal{U}(\rho)=U\rho U^\dagger$ for a Haar-random unitary $U$, while $\Phi$ is a maximally random CPTP map constructed from truncated Haar unitaries on an extended Hilbert space with $d$ dissipative channels. The Kraus rank of the resulting map is therefore $D=d+1$.

The parameter $p$ controls the deviation from unitarity, while the number of channels $d$ sets the typical strength of dissipative contraction. In this ensemble, the spectrum of $\mathcal{E}_p$ exhibits a geometric transition. For fixed $d$ and small $p$, the nontrivial eigenvalues populate a ring in the complex plane. Beyond a critical value $p_c(d)$, the inner radius collapses and the support becomes disk-like inside the unit circle~\citep{saSpectralTransitionsUniversal2020}.

From the RMT perspective, the dissipative part can be treated in an effective non-Hermitian random-matrix model, where $\mathcal{E}_p$ is replaced by a linear combination of a CUE matrix and a Ginibre matrix. This makes it possible to compute analytically the inner and outer radii, $R_-(p)$ and $R_+(p)$, of the spectral support in the large-$N$ limit using quaternionic free probability~\cite{saSpectralTransitionsUniversal2020}
\begin{equation}
\label{eq:radii_diluted_unitaries}
R_{\pm}(p)
=
\frac{1}{\sqrt{d}}\,
\sqrt{(1-p)^2 d \pm p^2}\, .
\end{equation}
The outer radius $R_+(p)$ decreases from $1$ at $p=0$, reaches a minimum below $1/\sqrt{d}$, and then increases back to $1/\sqrt{d}$ at $p=1$. The inner radius $R_-(p)$ decreases from $1$ at $p=0$ and vanishes at $p_c(d)$, beyond which the support is a disk. The relaxation gap, $-\log R_+(p)$, is therefore a non-monotonic function of $p$, showing that stronger nonunitarity does not necessarily imply faster mixing. In the language of DQC, this ring--disk transition provides a clean tunable spectral diagnostic of the crossover from almost unitary dynamics with long-lived modes to strongly mixing dynamics with broadly distributed decay rates.

The same parametric structure appears in the local Kraus circuits studied in Ref.~\citep{saSpectralTransitionsUniversal2020}. There, each time step consists of layers of two-site CPTP gates with the same mixture parameter $p$, and the spectrum of the full Floquet map again evolves from an annular support with weak dissipation to a disk-like support when the strength of dissipation increases. Finite system sizes smear the spectral boundaries and produce real-axis eigenvalues associated with real Ginibre statistics, but the transition remains clearly visible. This highlights that the ring--disk scenario is robust under locality and circuit structure, and that complex-spectrum diagnostics, such as complex spacing ratios (see Section~\ref{complex-spacing-ratios-csr}), must be interpreted with care depending on whether the system is in the annular, quasi-unitary regime or in the deeply dissipative, disk-like regime.

The ring-disk transition has also been detected experimentally, in a NISQ setting. In Ref.~\citep{woldSpectraNoisyParameterized2025}, parameterized circuits were implemented on an IBM quantum processor, and the resulting nonunitary dynamics was reconstructed as a CPTP map using a tailored tomography. Depending on circuit depth, the retrieved spectra form either a ring or a disk, and the ring--disk transition was detected upon the increase of the depth. Moreover, the measured spectra are quantitatively close to the diluted-unitary framework~\citep{saSpectralTransitionsUniversal2020}, thus making the ring-disk scenario a practical classifier of NISQ circuit implementations.

Beyond its role as a solvable DQC toy model, the diluted-unitary construction has become a useful framework for modeling local and global dissipation. In noisy random-circuit implementations, it provides a compact effective description of how gate and permutation errors accumulate and degrade performance in architectures ranging from fully connected to one-dimensional connectivity, with analytic control over fidelity decay and its scaling with system size and depth~\citep{samosFidelityDecayError2025}. More recently, the same idea of unitary dynamics diluted by a random Kraus component of controlled rank was used to quantify how integrability signatures persist at weak dissipation and disappear beyond a dissipation threshold, including spectral diagnostics such as angular velocity of the eigenvalue and universal estimates of the disappearance of clustering structures~\citep{pereiraDissipationInducedThresholdIntegrability2025}.

\subsubsection{Other models of random maps}\label{other-families-of-random-parametric-quantum-channels}

A complementary parametric route between coherent and dissipative
dynamics is provided by the \emph{random parametric quantum channels}
(RPQC) introduced in
Ref.~\citep{matsoukas-roubeasQuantumChaosCoherence2024}. In that work,
a two-parameter family of CPTP maps,
\begin{equation}
\Lambda_{\tau,\varepsilon}[\rho]
=
(1-\varepsilon)\,e^{-i\tau H/\hbar}\rho\,e^{+i\tau H/\hbar}
+
\varepsilon\sum_{r=1}^{K} N_r\rho N_r^\dagger ,
\end{equation}
was considered. Here $\varepsilon\in[0,1]$ sets the strength of dissipation, $\tau$ plays
the role of a stroboscopic ``dissipation period'', $H$ is sampled from
a random-matrix ensemble, and the Kraus operators $N_r$ are themselves
constructed from random matrices~
\citep{matsoukas-roubeasQuantumChaosCoherence2024}. This model is
motivated as a prototype of unitary evolution periodically interrupted
by measurements or transient environmental couplings, and in the
weak-noise/short-$\tau$ limit it admits a continuous-time Lindblad
description with Lindblad operators inherited from the Kraus operators
\citep{matsoukas-roubeasQuantumChaosCoherence2024}. This model connects discrete-time random maps with continuous-time random Lindbladian dynamics.

In the RPQC ensemble, the bulk of the spectrum exhibits a
rich phase diagram as a function of $(\tau,\varepsilon)$. For large
$\tau$, one recovers rotationally invariant ring and disk phases
similar to those found in random Kraus maps, with a ring--disk
transition as $\varepsilon$ is increased
\citep{saSpectralTransitionsUniversal2020,matsoukas-roubeasQuantumChaosCoherence2024}.
At smaller $\tau$, however, rotational invariance is broken and the
spectrum develops shifted disks and crescent-like shapes, reflecting the
competition between coherent rotation and dissipative contraction in the
complex plane~\citep{matsoukas-roubeasQuantumChaosCoherence2024}. This modifies the ring--disk scenario by showing that additional control parameters, such as the stroboscopic period, can constrain the angular structure of the spectrum even when the noise strength is fixed. Thus, the degree of nonunitarity cannot be inferred from the radial distribution alone.

\medskip

In summary, three key classes of random maps highlighted here -- (i)
global random quantum operations \`a la Bruzda \emph{et al.}~\citep{bruzdaRandomQuantumOperations2009}, (ii) parametric random Kraus maps and local circuits \`a la S\'a \emph{et al.}~\citep{saSpectralTransitionsUniversal2020}, and (iii) random parametric quantum channels \`a la Matsoukas-Roubeas \emph{et al.}~\citep{matsoukas-roubeasQuantumChaosCoherence2024}---provide a coherent ladder of models connecting abstract random-matrix constructions to realistic nonunitary circuits. They underpin many things highlighted in this chapter. Namely, global random channels offer analytically tractable benchmarks for spectral universality, parametric Kraus maps and circuits furnish a clean realization of the ring--disk transition and steady-state universality in DQC, and RPQC models show how static spectral chaos and dynamical coherence diagnostics can diverge under noise.

\subsubsection{Integrable quantum maps as a foil to
randomness}\label{integrable-quantum-maps-as-a-foil-to-randomness}

Alongside random map ensembles, integrable open quantum maps are important as controlled counterexamples to the random-matrix universality. Integrable open quantum circuits with CPTP step maps were constructed in Ref.~\cite{saIntegrableNonunitaryOpen2021} and later in Ref.~\cite{lei2022PRB}. These models are nonunitary but remain constrained by an underlying integrable structure. As a result, the spectra of the corresponding Lindbladians and maps and the dynamics they induce can be analyzed using integrability techniques, leading to non-generic eigenvalue correlations and spectral supports despite the presence of dissipation. From the DQC perspective, such integrable maps are valuable because they provide explicit control cases where complex spectra and steady states do not display Ginibre-like universality, and they help disentangle which ``chaos indicators'' are tied to non-integrability, such as complex-spacing-ratio statistics, from generic consequences of CPTP contraction, such as spectral radius $\leq 1$ and the existence of a fixed point.

%%%%%%%%%%%%%%%%%%%%%%%%%%%%%%%%%%%%%%%%%%%%%%%%%%%%%%%%%%%%%
\section{Spectral Statistics}
\label{spectral}

\subsection{Local Level Statistics}
\label{sec:local_stat}

Local spectral statistics---correlations between nearby
eigenvalues---are among the sharpest and most universal diagnostics of
quantum chaos. In closed (Hamiltonian) systems the spectrum lies on the
real axis, and local correlations are captured by the nearest--neighbour
spacing distribution $P(s)$, which famously distinguishes integrable
(Poisson statistics, level clustering) from chaotic dynamics
(Wigner--Dyson statistics, level repulsion). In practice, however,
$P(s)$ requires \emph{unfolding} (removing the nonuniversal energy
dependence of the mean level density)~\cite{Stockmann1999}, which can be ambiguous in 
many-body settings. A particularly robust alternative is to use ratios
of consecutive gaps, popularized in many-body contexts~\cite{oganesyanLocalizationInteractingFermions2007} because they are
\emph{unfolding-free}. The key object is 
\begin{equation}
r_n=\frac{s_{n}}{s_{n-1}},\qquad \tilde r_n=\frac{\min(s_n,s_{n-1})}{\max(s_n,s_{n-1})},
\end{equation}
with $s_n=E_{n+1}-E_n$. The ratio distribution admits accurate
Wigner-like surmises and provides clean reference values (e.g.~for
$\langle \tilde r\rangle$) for Poisson vs GOE/GUE/GSE universality
classes~\citep{atasDistributionRatioConsecutive2013}. This is now
a standard tool in many-body spectral studies, including early applications to
interacting localization problems where unfolding is particularly
delicate~\citep{oganesyanLocalizationInteractingFermions2007}.

For open systems, the operator controlling the dynamics---a Lindbladian, an effective non-Hermitian Hamiltonian, or a CPTP map---typically has a complex spectrum. Eigenvalues populate
a two-dimensional domain and repel each other in the plane, not on a line.
This makes ``local'' statistics geometrical, so we have to define nearest
neighbours in $\mathbb C$, and the unfolding problem becomes
significantly worse because there is no unique scalar map that flattens
an arbitrary 2D spectral density. These difficulties motivate
observables that (i) depend only on local geometry and (ii) avoid explicit unfolding.

\subsubsection{``Local'' statistics in the complex plane}\label{local-statistics-in-the-complex-plane}

Given complex eigenvalues $\lambda_k$, local statistics focus on the
arrangement of eigenvalues within a small neighbourhood of a reference
point $\lambda_k$. The most direct analog of the Hermitian spacing $s$
is the Euclidean distance to the nearest neighbour in $\mathbb C$, $
s_k = \min_{j\neq k}|\lambda_j-\lambda_k|
$. This metric was used in the work by Grobe, Haake, and Sommers~\cite{grobeQuantumDistinctionRegular1988}; see  Sections~\ref{history} (year 1988) and \ref{grobe_revision}.
They demonstrated that short-distance repulsion in the plane can distinguish regimes that are regular and chaotic in the semiclassical limit. A complementary (and historically important) development came from
lattice quantum chromodynamics (QCD) at finite chemical potential, where unfolding prescriptions
for complex spectra were explicitly constructed and tested in order to
extract nearest-neighbour spacing statistics in the plane~
\citep{markumNonHermitianRandomMatrix1999}.

However, both density inhomogeneity and spectral edges (``bulk vs
edge'') can strongly bias distance-based statistics unless unfolding is handled with care.

\begin{figure}
    \centering
    \includegraphics[width=1\linewidth]{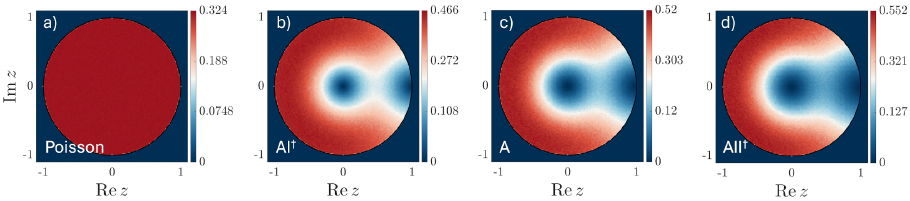}
    \caption{Distributions of complex spacing ratios, Eq.~\eqref{eq:csr}, for different universality classes. 
Panel (a) shows the CSR probability distribution for the two-dimensional Poisson point process, corresponding to uncorrelated complex eigenvalues. 
Panels (b)--(d) show the random-matrix predictions for the three non-Hermitian bulk classes 
AI$^\dagger$, A, and AII$^\dagger$, respectively. 
The depletion near $z=0$ and the anisotropic structure reflect eigenvalue repulsion in the complex plane.
}
    \label{fig:CSR}
\end{figure}

\subsubsection{Complex spacing ratios (CSR)}\label{complex-spacing-ratios-csr}

The central idea of Ref. \citep{saComplexSpacingRatios2020a} is to
generalize spacing ratios to complex spectra by using nearest neighbour
(NN) and next-to-nearest neighbour (NNN) distances defined geometrically
in the plane. For each eigenvalue $\lambda_k$, let
$\lambda^{\text{NN}}_k$ and $\lambda^{\text{NNN}}_k$ denote the
closest and second-closest eigenvalues in $\mathbb C$. The complex
spacing ratio is then defined as 
\begin{equation}
\label{eq:csr}
z_k=\frac{\lambda^{\text{NN}}_k-\lambda_k}{\lambda^{\text{NNN}}_k-\lambda_k}=re^{i\theta},
\end{equation}
which lives inside the unit disk. This
construction (i) needs no global ordering of eigenvalues, (ii) is
insensitive to local rescalings of the density, and (iii) retains
angular information, which turns out to be an exceptionally sensitive
marker of non-Hermitian correlations
\cite{saComplexSpacingRatios2020a}.

Two universal categories emerge:

\begin{itemize}
\item
  Uncorrelated (integrable/regular) complex spectra: 2D Poisson. For
  independent levels in the plane, the CSR distribution is flat
  inside the unit disk, i.e.~isotropic and constant. Equivalently, the
  angular marginal is uniform and the radial marginal is trivial
  ($\propto r$) \citep{saComplexSpacingRatios2020a}. This is the
  natural extension of Poisson statistics to $\mathbb C$.
\item
  Chaotic non-Hermitian spectra: Ginibre-type correlations. For
  non-Hermitian random matrices in the Ginibre universality class,
  eigenvalues exhibit strong short-range correlations: level repulsion
  in the plane and pronounced angular anisotropy in the CSR distribution
  \citep{saComplexSpacingRatios2020a}. CSR captures not only radial
  repulsion (suppression of small $r$) but also directional structure,
  which often provides the cleanest separation between Poisson and
  chaotic regimes.
\end{itemize}

This provides an open-system analogue of the ``Poisson vs
Wigner--Dyson'' dichotomy of closed systems, extending earlier
observations in dissipative maps
\citep{grobeQuantumDistinctionRegular1988} and aligning with systematic
integrability-to-chaos studies for dissipative generators
\citep{akemannUniversalSignatureIntegrability2019}. The CSR distributions for different Ginibre-type non-Hermitian bulk classes~\cite{saComplexSpacingRatios2020a} AI$^\dagger$, A, and AII$^\dagger$, corresponding to different symmetry classes as discussed in Section~\ref{classification}, are shown in Fig.~\ref{fig:CSR}. They are contrasted with the flat CSR distribution of the two-dimensional Poisson point process, corresponding to uncorrelated complex eigenvalues.

\subsubsection{Surmises and practical
implementation}\label{surmises-and-practical-implementation}

A subtlety emphasized in the CSR framework is that finite-size spectra can be strongly distorted by boundary effects. The cure is to benchmark against ensembles that reproduce the correct large-$N$ local physics while suppressing edge artifacts, leading to Wigner-like surmises that converge rapidly for both Hermitian-type and Ginibre-type universality
classes~\citep{saComplexSpacingRatios2020a}.

Operationally, CSR analysis typically proceeds as follows:

\begin{enumerate}
\def\labelenumi{\arabic{enumi}.}
\item
  Resolve symmetries and sectors. Superposing independent symmetry
  sectors can wash out correlations; this is particularly important for
  Lindbladians, where conserved quantities and antiunitary symmetries
  can impose nontrivial spectral structures
  \citep{sa2023PRX}.
\item
  Use both radial and angular information. In many dissipative settings
  the angular marginal (or simple angular moments) is more sensitive
  than radial moments alone \citep{saComplexSpacingRatios2020a}.
\end{enumerate}

\subsubsection{Follow-ups and representative uses of CSR}\label{follow-ups-and-representative-uses-of-csr}

Since its introduction, CSR has been adopted and extended in several directions:

\begin{itemize}
\item
  \textbf{Analytic progress for Ginibre CSR.} Dusa and Wettig~\cite{dusaApproximationFormulaComplex2022} derived an accurate
  approximation formula for CSR statistics in the Ginibre ensemble,
  providing useful closed expressions and moments for quantitative
  benchmarking \citep{dusaApproximationFormulaComplex2022,akemann2026arXiv}.
\item
  \textbf{Integrability--chaos transitions in dissipative generators.}
 Complex CSR-style statistics have been used to map the crossovers between  integrable and chaotic regimes in open quantum systems~
  \citep{akemannUniversalSignatureIntegrability2019}. More recently,
  Akemann \emph{et al.} identified two distinct
  transitions---Hermiticity breaking and integrability breaking---in
  complex eigenvalue statistics, emphasizing that different notions of
  ``non-Hermitian complexity'' can turn on at different parameter scales
  \citep{akemannTwoTransitionsComplex2025}.
\item
  \textbf{Many-body Lindbladians and symmetry classification.} CSR plays
  a natural role in sector-resolved studies of interacting Lindbladians,
  where the correct identification of the relevant non-Hermitian
  universality class requires symmetry resolution beyond the Hermitian
  Wigner--Dyson paradigm \citep{sa2023PRX} (see Sec.~\ref{classification}).
\item
  \textbf{Disordered interacting non-Hermitian systems and open
  boundaries.} CSR and related local probes have been used to
  characterize spectral correlations in disordered interacting
  non-Hermitian systems \citep{ghoshSpectralPropertiesDisordered2022a}
  and in non-Hermitian many-body localization settings with open
  boundaries \citep{sutharNonHermitianManybodyLocalization2022}.
\item
  \textbf{Integrability to chaos directly in Liouvillian families.} In
  structured families of quantum Liouvillians, CSR-type local
  diagnostics provide an unfolding-free way to chart integrable vs
  chaotic regions in parameter space
  \citep{rubio-garciaIntegrabilityChaosQuantum2022,wang2026arXiv}.
\end{itemize}

\subsubsection{Experimental detection}

The CSR diagnostic has  already been implemented experimentally in Ref.~\citep{Wold2025ExperimentalDetection}. The experiment realized dissipative quantum maps on a superconducting quantum processor by using a five-qubit circuit, with four system qubits and one environment qubit. After the circuit was executed, the environment qubit was traced out, producing an effective CPTP map acting on the four-qubit system.

Two circuit families were compared. The integrable family was a free-fermion, or matchgate, circuit~\cite{saIntegrableNonunitaryOpen2021,lei2022PRB} built from layers of single-qubit $Z$ rotations and $\sqrt{i\mathrm{SWAP}}$ gates arranged in a brickwork pattern. Integrability was broken by adding extra single-qubit $Y$ rotations in each layer, producing the chaotic circuit family. Thus the experimental setting directly realizes the same conceptual comparison used throughout this section, an integrable dissipative map with uncorrelated complex eigenvalues versus a chaotic dissipative map with Ginibre-type eigenvalue correlations.

The maps were reconstructed using gradient-based quantum process tomography~\citep{woldSpectraNoisyParameterized2025}, which enables the reconstruction of a quantum map in Kraus form from collected Pauli-string measurements. The tomography included a model of state-preparation and measurement errors, and the retrieved maps were then diagonalized to obtain their eigenvalues and the corresponding CSR distribution $P(z)$. 
%The processor had average relaxation time of order $150\,\mu\mathrm{s}$, single-qubit gate fidelity about $99.98\%$, and CZ-gate fidelity about $99.86\%$. 
The implemented circuits were transpiled into single-qubit gates and CZ gates and Pauli twirling was used to reduce unitary errors. 
%For each Pauli mode, $2000$ shots were collected for each of $6$ random twirling realizations, giving $12000$ shots in total. The reconstruction used full-rank Kraus maps and was optimized for $3000$ gradient steps with learning rate $10^{-3}$. Bootstrap resampling was then used to estimate the uncertainty of the CSR statistics.

For shallow circuits, in particular at depth $T=5$, the approximate particle-number symmetry of the free-fermion circuit is still visible. Therefore, the CSR statistics were evaluated in the steady-state, or half-filling, sector, in agreement with the general requirement (see Section~\ref{surmises-and-practical-implementation} )that independent symmetry sectors should not be mixed when testing spectral correlations. In the integrable circuit, the CSR distribution shows no depletion near $z=0$, consistently with the absence of level repulsion. In contrast, the chaotic circuit displays the characteristic hole around the origin, as expected for correlated complex spectra in the Ginibre universality class. This provides an experimental realization of the distinction between Poisson-like and Ginibre-like CSR statistics discussed above.

\begin{figure}
    \centering
    \includegraphics[width=0.98\linewidth]{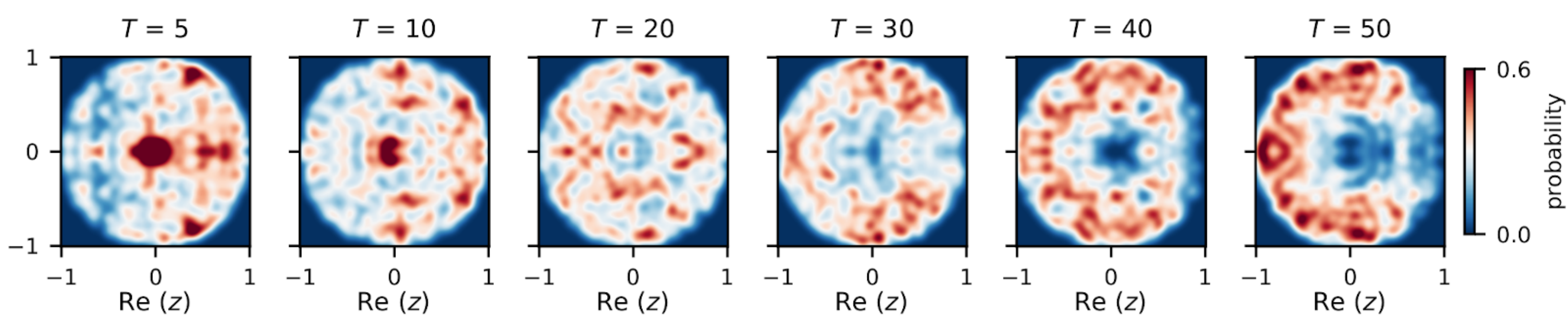}
    \caption{
   Complex spacing ratio distributions retrieved for four-qubit dissipative quantum maps obtained from integrable free-fermion circuits implemented on a superconducting quantum processor~\citep{Wold2025ExperimentalDetection}. The panels show the CSR density $P(z)$ for increasing circuit depth $T$. For small $T$, the distribution retains integrable features, including a peak near $z=0$ and the absence of level repulsion. With increasing depth, accumulated hardware noise breaks the effective integrable structure and the distribution crosses over toward the depleted profile characteristic of chaotic dissipative dynamics. %Adapted from Ref.~\citep{Wold2025ExperimentalDetection}.
    }
    \label{fig:Figure11}
\end{figure}

Figure~\ref{fig:Figure11} shows the effect of increasing circuit depth in the experimentally implemented integrable circuit. 
Although the ideal free-fermion circuit is integrable, the accumulated hardware noise acts as an additional dissipative chaotic component. 
As $T$ is increased from $5$ to $50$, the CSR distribution evolves from an integrable-like profile with a pronounced peak near the origin (typical of low-dimensional free-fermion maps~\citep{Wold2025ExperimentalDetection}) toward a depleted, annular structure close to the CSR shape typical of chaotic AI-class; see Figure~\ref{fig:CSR}. %The same crossover was studied for different couplings between the last system qubit and the environment qubit, using $\sqrt{i\mathrm{SWAP}}$ for stronger coupling and $\sqrt[6]{i\mathrm{SWAP}}$ for weaker coupling. In both cases the large-depth behavior approaches the chaotic finite-size benchmark, showing that intrinsic device noise itself can drive the observed transition to dissipative quantum chaos.

%%%%%%%%%%%%%%%%%%%%%%%%%%%%%%%%%%%
\subsection{Spectral form factors}
\label{sec:SFF}

The spectra of complex quantum systems are not only correlated at short distances, as measured, e.g., by spacing ratios. The simplest quantity capturing universal correlations at all scales is the two-point correlation function $R_{2}(E_1-E_2)$ of two eigenvalues $E_1$ and $E_2$. In Hermitian systems, it is often more convenient to work with its Fourier transform, also known as the spectral form factor (SFF)~\cite{mehta2004,leviandier1986,brezin1997,prange1997}:
\begin{equation}
	\label{eq:def_SFF}
	K(t)=\left\langle\left|\Tr U_t\right|^2\right\rangle=
	\left\langle\sum_{j,k}e^{-i\left(E_j-E_k\right)t}\right\rangle
	=\int\d E_1 \d E_2 \, R_{2}(E_1-E_2)\, e^{-i(E_1-E_2) t},
\end{equation}
The first equality shows that the SFF is also directly related to the dynamics through the time evolution operator $U_t=\exp\{-i H t\}$.
The SFF can be computed analytically in RMT~\cite{mehta2004,guhr1998} and displays a characteristic dip-ramp-plateau structure (in contrast, for Poisson spectra characteristic of integrable systems, the SFF decays directly to the plateau). For times longer than the Heisenberg time proportional to the inverse level spacing, the SFF is flat and reflects the discreteness of the spectrum. At earlier times, the spectral rigidity of chaotic spectra gives a ramp in the SFF whose analytic form is solely determined by symmetry. For sufficiently short times, level statistics of realistic quantum chaotic Hamiltonians deviate from the RMT prediction. Determining the timescale for the onset of deviations, the so-called Thouless time, in many-body systems is a topic of current interest~\cite{garcia-garcia2016PRD,garcia-garcia2018PRD,kos2018PRX,gharibyan2018JHEP,chan2018PRL,yoshimura2024}.

In dissipative systems, the spectral and dynamical definitions of the SFF no longer coincide. Two different (albeit similarly named) non-Hermitian versions of the SFF were defined that focus on these two different aspects.
The dissipative form factor (DFF), introduced in Ref.~\cite{can2019JPhysA}, generalizes the dynamical definition of the SFF and is given by the trace of the quantum map $\Phi_t$,
\begin{equation}
    \mathrm{DFF}(t)=
    \left\langle\Tr\Phi_t\right\rangle=
    \left\langle \sum_{\mu_1\cdots \mu_t}\left|\Tr K_{\mu_1}\cdots K_{\mu_t}\right|^2\right\rangle=
    \left\langle\sum_j \lambda_j^t\right\rangle.
\end{equation}
The late-time behavior of the DFF is controlled by the spectral gap; it was used to compute the latter for random Liouvillians~\cite{can2019JPhysA}, dynamical phase transitions in the Lindbladian SYK model~\cite{kawabata2023PRB}, and evaluate the robustness of quantum chaos to weak dissipation and establish a link to irreversibility in isolated systems~\cite{yoshimura2024,yoshimura2025}.

Contrary to the SFF, the DFF does not measure the correlations of eigenvalues in the complex plane, which are instead captured by the dissipative spectral form factor (DSFF), defined in Ref.~\cite{li2021PRL} (see also Refs.~\cite{Braun2001,fyodorov1997}) as the complex Fourier transform of the two-point correlator of the eigenvalues of $\mathcal{L}$ (or its exponentiated version $\Phi$),
\begin{equation}
    \mathrm{DSFF}(\tau)=
    \left\langle \sum_{j,k}e^{-i \left\{\Re(\ell_j-\ell_k)\Re \tau + \Im(\ell_j-\ell_k)\Im\tau\right\}}\right\rangle
    =\int\d^2 z_1 \d^2 z_2 \, R_{2}(z_1-z_2)\, e^{-i\left\{\Re (z_1^*\tau)-\Re (z_2^* \tau)\right\}/2}.
\end{equation}
If the ``complex time'' (i.e., the Fourier variable) is given by $\tau=t \exp{i \theta}$, then the DSFF reduces to the standard SFF (in time $t$) of the projection of the eigenvalues along the ray in the complex plane defined by $\theta$~\cite{li2021PRL,garcia2023PRD}. It was computed analytically for class A in Refs.~\cite{li2021PRL,fyodorov1997} and numerically for classes AI$^\dagger$ and AII$^\dagger$ in Ref.~\cite{garcia2023PRD}. The DSFF of non-Hermitian random matrices still has a dip-ramp-plateau structure, but the ramp is no longer linear. Moreover, the Heisenberg time scales with $\sqrt{N}$ instead of $N$, where $N$ is the Hilbert space dimension. The DSFF accurately captures spectral correlations of several single- and many-body physical systems~\cite{ghoshSpectralPropertiesDisordered2022a,li2021PRL,garcia2023PRD,chan2022NatComm,shivam2023PRL,sen2025}, while its limitations (e.g., nonstationarity and the potential absence of a correlation hole) were highlighted in Ref.~\cite{garcia2023PRD}.

Yet a third non-Hermitian generalization of the SFF involves identifying the latter as the survival probability of a coherent Gibbs state under the unitary evolution $U_t$; this can then be straightforwardly extended to nonunitary evolution, as in Refs.~\cite{matsoukas-roubeasQuantumChaosCoherence2024,xu2021PRB,cornelius2022PRL,matsoukas-roubeas2023PRA,gopalakrishnan2026arXiv} (see also Refs.~\cite{delcampo2020JHEP,matsoukas-roubeas2023JHEP}). It is a convenient tool to study the effect of decoherence on Hamiltonian chaos, which, for the particular case of dissipators that preserve the energy eigenbasis, can be either suppressed~\cite{xu2021PRB} or enhanced~\cite{cornelius2022PRL}. We note that this non-Hermitian form factor is closely related to the DFF; indeed, the latter can also be obtained from averaging the survival probability over Haar-random initial states~\cite{yoshimura2024,kawabata2023PRB}.

%%%%%%%%%%%%%%%%%%%%%%%%%%%%%%%%%%%%%%%%%%%%%%%%%%%%%%%%%%%%%
\section{Symmetry Classifications}
\label{classification}

The interplay of symmetries, correlations, and dynamics lies at the heart of our understanding of complex interacting quantum many-body systems. It provides a compact and powerful framework for obtaining universal information not otherwise available for generic quantum systems.
Hermitian Hamiltonians are classified by a small set of unitary and antiunitary symmetries~\cite{dyson1962a}. The behavior under time-reversal, particle-hole, and chiral symmetry places them in one of ten classes, the Altland-Zirnbauer classes~\cite{altland1997}. In turn, the quantum chaos conjecture~\cite{bohigas1984} states that, if the system is chaotic, the Hamiltonian displays the statistical behavior of a random matrix from the same symmetry class. Finally, the correlations of random matrices are universal and fixed by symmetry. That is, level repulsion measures the behavior under time reversal, while the spectral density close to the origin is determined by particle-hole and chiral symmetry.
Thus, certain observables like conductance fluctuations in disordered electronic systems follow directly from symmetry transformations.

\subsection{Non-Hermitian matrices}

As discussed above, open quantum systems are modelled by non-Hermitian matrices with complex spectra and their classification is far more involved. 

The classification of non-Hermitian matrices in terms of their global symmetries was first attempted in Refs.~\cite{bernard2002,magnea2008}, after it was realized that non-Hermitian matrices admit more possible antiunitary symmetries than Hermitian ones (because complex conjugation and transposition become distinct transformations). Years later, the subject was revisited~\cite{sato2012PTP,lieu2018PRB,gong2018PRX,kawabata2019NatComm} and the role of different antiunitary symmetries was clarified, finally culminating in a 38-fold classification for matrices with a point-gap spectrum~\cite{kawabata2019PRX,zhou2019PRB} and a 54-fold classification for matrices with a line-gap spectrum~\cite{liu2019PRB,ashida2020}. These classifications have become known as the Bernard-LeClair (BL) classification. As is always the case in symmetry classifications, it is assumed that the non-Hermitian Hamiltonian is irreducible, i.e., that it cannot be decomposed into independent blocks due to a commutation with a unitary symmetry. The Hamiltonian is then classified by the action of four antiunitary operators ($T_\pm$ and $C_\pm$) and three types of unitary operators ($P$ and $Q_\pm$):
\begin{equation}
	\label{eq:sym_def}
	\begin{split}
	& T_\pm H T_\pm^{-1} = \pm H, \quad T_\pm i T_\pm^{-1}=-i,\quad T_\pm^2=\pm1,\\
	& C_\pm H^\dagger C_\pm^{-1} = \pm H, \quad C_\pm i C_\pm^{-1}=-i,\quad C_\pm^2=\pm1. \\
	& P H P^{-1} = \pm H, \quad P i P^{-1}=i,\quad P^2=+1,\\
	& Q_\pm H^\dagger Q_\pm^{-1} = \pm H, \quad Q_\pm i Q_\pm^{-1}=i,\quad Q_\pm^2=+1.
	\end{split}
\end{equation}
There can exist at most one operator of each type (otherwise, their product would be a commuting unitary symmetry which we assumed not to exist); we denote the absence of a symmetry as it squaring to zero. As was shown in Ref.~\cite{sathesis}, when at least one antiunitary symmetry exists, then the non-Hermitian symmetry class is completely specified by the squares of the four operators. The combined action of two antiunitary operators of different types gives unitary involutive transformations, which therefore only label new classes if no antiunitary symmetries exist; these classes are again fully labeled by the squares of the three unitary involutions. After identifying Hamiltonians differing by a global phase, counting the possible combinations yields 38 symmetry classes~\cite{kawabata2019PRX,zhou2019PRB,sathesis} (these works used different ways of labeling the classes, but they can all be shown to be equivalent).

The BL classification has been most influential in investigations of non-Hermitian single-particle physics along the lines of the Hatano-Nelson model.
In the many-body setting, the non-Hermitian SYK model~\cite{garcia2022PRX} and its chiral extensions~\cite{sa2022PRD} provided concrete realizations of 19 of the 38 symmetry classes, depending on the values of $q$ and $N$.

\subsection{Lindbladians}

Because non-Hermitian Hamiltonians provide only an effective description of open quantum dynamics (for short evolution time or by postselecting jump-free quantum trajectories), a fundamental question was how the classification is constrained by the stringent constraints of Lindbladian dynamics--- Hermiticity- and trace-preservation and complete positivity. Using causality arguments, Ref.~\cite{lieu2020PRL} argued that there are ten classes of single-particle Lindbladians. However, they did not consider shifting the spectral origin~\cite{prosen2012PRL,prosen2012PRA}, which avoids the causality restrictions, as pointed out in Ref.~\cite{kawasaki2022PRB}. Allowing for this possibility, Ref.~\cite{kawasaki2022PRB} showed that all BL classes can be implemented at the level of single-particle Lindbladians. Ref.~\cite{altland2021PRX} considered the invariance of the dynamics under linear or antilinear, canonical or anticanonical transformations of fermionic creation and annihilation operators, leading to a tenfold classification of the single-particle and interactions tensors that appear in the Hamiltonian and jump operators. A complete many-body matrix classification of Lindbladians was performed in Refs.~\cite{sa2023PRX,kawabata2023PRXQ}. 

Representing the Lindbladian operator as a matrix in a doubled Hilbert space and performing a necessary spectral shift, the symmetries of Eq.~\eqref{eq:sym_def} are promoted to (super)operator symmetries \cite{sa2023PRX}. It was shown that, for irreducible Lindbladians (i.e., those without any commuting unitary symmetries) Hermiticity-preservation implies that a $T_+^2=1$ symmetry always exists and is given by the product of the swap operator that exchanges the two copies of the Hilbert space (bra and ket) and complex conjugation. Moreover, it follows from the two-copy structure that $C_+^2=-1$ symmetries are not allowed, with the consequence that no classes with Kramers' degeneracy exist in Lindbladian dynamics. On the other hand, trace-preservation and complete positivity do not constrain the allowed symmetry classes; the above discussion can be extended to non-trace-preserving and even non-Markovian settings, as long as they are described by a local-in-time master equation~\cite{sa2023PRX}.
Careful bookkeeping leads to a tenfold classification, which can, however, be enriched by the presence of commuting unitary symmetries. Indeed, in such a situation, after block-reducing the Lindbladian, it may happen that the $T_+^2=+1$ connects different blocks, that is, although the full Lindbladian must preserve Hermiticity, individual blocks must not (sectors containing steady states always do, and hence belong to the simpler tenfold classification). In this case, there are 19 additional allowed symmetry classes (note that the precise number depends on whether one considers the 38-\cite{kawabata2023PRXQ,garcia2024PRD} or 54-fold \cite{sa2023PRX} classification as basis for the Lindbladian classification; the additional 19 classes correspond to the latter). Ref.~\cite{sa2023PRX} constructed simple, physically relevant examples of dephasing and driven spin chains in all classes of the tenfold classification and an example that goes beyond it. 

The Lindbladian symmetry classification was also extensively tested in the Lindbladian SYK model~\cite{kawabata2023PRXQ,garcia2024PRD}, as tuning the parameters of the model, namely the number of Majoranas $N$, the complexity of the interactions $q$, and the complexity of the jump operators $r$ allows one to cycle through 11 distinct classes (A, AI, AIII, BDI, CI, AI$^\dagger$,  BDI$^\dagger$, CI$^\dagger$, BDI$_{++}$, BDI$_{-+}$, CI$_{+-}$, and CI$_{--}$)~\cite{garcia2024PRD}. In particular, it includes 8 out of the 10 classes of the tenfold classification steady-state sectors (i.e., without unitary symmetries).

The preceding discussion extends beyond Lindbladian dynamics to any operator that has an antiunitary two-copy swap symmetry. As an example, Ref.~\cite{garcia2024PRD} provided a classification of fermionic PT-symmetric Hamiltonians, consisting of two quantum systems that are the time-reversal image of each other, coupled by a swap-invariant interaction. Finally, a two-copy structure with antiunitary swap symmetry prevents Kramer's degeneracy even in Hermitian settings~\cite{garcia2024PRL}.

\subsection{Spectral consequences}

The existence of antiunitary symmetries of a non-Hermitian Hamiltonian $H$ (or Lindbladian) constrains both its eigenvalues and eigenvectors (the action of unitary involutions can be obtained by combining antiunitary transformations). As shown in Fig.~\ref{fig:CSR}, $C_+$ determines the level statistics of the bulk eigenvalues of $H$, and is most conveniently captured by CSRs.
On the other hand, $T_+$, $T_-$, and $C_-$ symmetries introduce nonlocal correlations between eigenvalues and eigenvectors. They reflect the spectrum across the real axis, the imaginary axis, and the origin, respectively, while also connecting the respective right and left eigenvectors. The existence of any two such symmetries implies the third; in that case, the spectrum has dihedral symmetry~\cite{sa2023PRX,prosen2012PRL}.
More concretely, if $|\phi_\lambda\rangle$ and $|\tilde{\phi}_\lambda\rangle$ are respectively the right and left eigenvectors of $H$ with eigenvalue $\lambda$, i.e., $H|\phi_\lambda\rangle=\lambda|\phi_\lambda\rangle$ and $H^\dagger|\tilde{\phi}_\lambda\rangle=\lambda^*|\tilde{\phi}_\lambda\rangle$, then we have the identities:
\begin{equation}
	\label{eq:spectral_consequences}
\begin{split}
	&H(T_+|\phi_\lambda\rangle)=\lambda^* (T_+|\phi_\lambda\rangle)
	\implies T_+|\phi_\lambda\rangle=|\phi_{\lambda^*}\rangle,
	\\
	&H(T_-|\phi_\lambda\rangle)=-\lambda^* (T_-|\phi_\lambda\rangle)
	\implies T_-|\phi_\lambda\rangle=|\phi_{-\lambda^*}\rangle,
	\\
	&H^\dagger(C_+|\phi_\lambda\rangle)=\lambda^* (C_+|\phi_\lambda\rangle)
	\implies C_+|\phi_\lambda\rangle=|\tilde{\phi}_{\lambda}\rangle,
	\\
	&H^\dagger(C_-|\phi_\lambda\rangle)=-\lambda^* (C_-|\phi_\lambda\rangle)
	\implies C_-|\phi_\lambda\rangle=|\tilde{\phi}_{-\lambda}\rangle.
\end{split}
\end{equation}
The first line tells us that a $T_+$ symmetry implies that $\lambda^*$ is also an eigenvalue of $H$ and that the respective right eigenvectors are related by $T_+$ (and similarly for the left eigenvectors). Similarly, $T_-$ forces $-\lambda^*$ to be an eigenvalue and again pairs the right eigenvectors of the two eigenvalues. On the other hand, $C_-$ connects the right eigenvector of $\lambda$ with the left eigenvector of $-\lambda$ (and vice versa). Finally, $C_+$ does not fix the existence of additional eigenvalues, but relates the right and left eigenvectors of every eigenvalue.

While CSRs can only distinguish three universality classes of bulk correlations determined by $C_+$, the level statistics on the axis of spectral reflection (in the case of $T_\pm$) or the spectral origin ($C_-$) are modified by the presence of such symmetries (and determined by the square of the symmetry). For $T_+$, the reflection symmetry changes the statistics near the real axis and induces the existence of purely real eigenvalues with their own local statistics; these effects have been studied for several different symmetry classes~\cite{garcia2024PRD,lehmann1991PRL,kanzieper2005PRL,forrester2007PRL,xiao2022}. Similar considerations hold for $T_-$ by noting that the respective spectra are related by a rotation of $\pi/2$. On the other hand, $C_-$ changes the statistics of the eigenvalues near the origin~\cite{garcia2022PRX,garcia2024PRD,akemann2009JPhysA,akemann2009PRE}. Strikingly, in the so-called microscopic regime of up to a few level spacings away from the spectral origin, not only are universal local correlations distinct from the bulk, but even the spectral density is universal and determined only by the symmetry~\cite{garcia2022PRX,garcia2024PRD,garcia2025PRB}.

Finally, the eigenvectors themselves have statistical properties dictated by symmetry. These are best seen in the Chalker-Mehlig eigenvector overlap matrix $O_{\alpha\beta}=\langle \tilde{\phi}_\alpha|\tilde{\phi}_\beta\rangle \langle \phi_\beta|\phi_\alpha\rangle$~\cite{mehlig1998PRL,mehlig2000JMP}.  The distribution of the entries and the first moments have been intensely investigated for the Ginibre ensembles~\cite{mehlig1998PRL,mehlig2000JMP,janik1999PRE,fyodorov2018,bourgade2020,akemann2020Acta}. The distribution of the diagonal overlaps $O_{\alpha\alpha}$ depends only on the value of $C_+^2$. On the other hand, using the constraints of Eq.~(\ref{eq:spectral_consequences}), it was shown in Ref.~\cite{sa2023PRX} that (i) the sign of the overlaps of eigenvectors connected by $C_-$ is given by $C_-^2$; and (ii) if $T_-^2=-1$, then the overlaps between eigenvectors connected by $T_-$ vanish. The statistical properties of the eigenvectors have been verified in spin-chain models~\cite{sa2023PRX} and the PT-symmetric SYK model~\cite{garcia2024PRD}.

\section{Semiclassical Signatures of DQC}
\label{semi}

\subsection{Semiclassical Limits and Classical Phase Space}\label{semiclassical-limits-and-classical-phase-space}

A semiclassical limit requires a tunable large parameter, typically a spin size \(S\), a particle number \(N\), or a large mode occupation, such that an effective Planck constant, for example \(\hbar_{\mathrm{eff}}\sim 1/S\) or \(1/N\), becomes small. In this limit, expectation values of collective observables can close into nonlinear mean-field equations, and the open quantum dynamics acquires an associated classical dissipative flow. The relevant phase-space structures are therefore not only the invariant tori and chaotic seas familiar from Hamiltonian semiclassics, but also fixed points, limit cycles, multistable basins, bifurcations, and strange attractors.

This mapping is made possible by the existence of quantum states that are already classical-like. Bosonic coherent states, spin coherent states, and more general group-theoretic coherent states form overcomplete families labeled by points of a classical phase space. They minimize fluctuations in an appropriate sense and allow quantum states to be represented by phase-space distributions such as Wigner functions, Husimi functions, or stochastic coherent-state mixtures. In large-\(S\), large-\(N\), or large-occupation limits, these distributions can become sharply localized, so that the motion of their centroids follows the mean-field equations while their finite width encodes quantum fluctuations. This is the basic reason why one can compare an open quantum model with a classical dissipative flow rather than merely using classical language as an analogy \cite{breuerTheoryOpenQuantum2010,gardinerQuantumNoiseHandbook2010}.

The semiclassical program in DQC asks what survives when this classical-like picture is quantized. Strict mean-field dynamics gives deterministic equations for expectation values, while the next corrections introduce noise, diffusion, or finite-width effects through Langevin equations, truncated-Wigner dynamics, or stochastic coherent-state decompositions. The central question is then where the dissipative phase-space structure appears in the quantum theory: in phase-space distributions, in the Liouvillian spectrum, in the steady-state density matrix, in reduced observables, or in individual quantum trajectories \cite{rufoQuantumSemiClassicalSignatures2025,duttaQuantumOriginLimit2025,mondalDissipativeChaosSteady2025}.

% Historically, the kicked-top analysis of Grobe, Haake, and Sommers provided the first sharp version of this question. They generalized spectral-statistics ideas to dissipative quantum maps and argued that regular and chaotic classical dissipative motion could be distinguished by correlations among complex eigenvalues, with cubic level repulsion emerging as a hallmark of the chaotic regime \cite{grobeQuantumDistinctionRegular1988}. This established an influential semiclassical correspondence principle for DQC: chaotic classical attractors should leave random-matrix-like signatures in the spectrum of the open quantum propagator. Later work has shown that this correspondence is useful but not universal. In the open Dicke model and in the dissipative kicked top, Villasenor and collaborators found Ginibre-type spectral correlations even when the classical dynamics has only simple attractors, and also regimes where chaotic classical motion does not produce the expected spectral signature \cite{villasenorAnalysisChaosRegularity2024a,villasenorBreakdownQuantumDistinction2024,villasenorCorrespondencePrincipleDissipation2025}. Thus, the semiclassical viewpoint remains central, but it cannot be reduced to a one-to-one identification between a classical attractor and a single quantum spectral diagnostic.

\subsection{Routes to Chaos in Open Models}\label{routes-to-chaos-in-open-models}

Some of the clearest successes of the semiclassical approach come from concrete open models where the route to dissipative chaos can be followed in detail. In these systems, the onset of DQC is not a featureless crossover. It is often mediated by standard mechanisms of nonlinear dynamics, including Hopf bifurcations, period-doubling cascades, Shilnikov scenarios, homoclinic collisions, crises, and the coexistence of multiple attractors.

Optomechanical systems provide a simple example. Bakemeier, Alvermann, and Fehske identified a period-doubling route to chaos and related the resulting nonlinear dynamics to observable changes in the optical spectrum \cite{bakemeierRouteChaosOptomechanics2015}. Collective light-matter systems provide another important family. In the unbalanced open Dicke model, Stitely et al.~showed that Hopf bifurcations generate superradiant oscillations, successive period doublings create symmetry-related chaotic attractors, and homoclinic bifurcations can merge those attractors into a larger chaotic set \cite{stitelyNonlinearSemiclassicalDynamics2020}. The dissipative anisotropic two-photon Dicke model adds further structure: Li, Fazio, and Chesi found limit cycles born through Hopf bifurcations, a pole-flip transition induced by anisotropy, and chaotic regions generated by period-doubling cascades in which symmetric attractors collide and fragment \cite{liNonlinearDynamicsDissipative2022}.

Increasing the number of modes enriches the same phenomenology. For the open Dicke trimer, Zhang et al.~showed that the bifurcation structure extends to a three-mode setting with additional superradiant phases and a more elaborate coexistence landscape \cite{zhangClosedOpenUnbalanced2024}. Driven-dissipative Bose-Hubbard dimers provide one of the most thoroughly analyzed cases. Giraldo and collaborators mapped pitchfork, Hopf, Shilnikov, flip, and T-point mechanisms, showing that dissipative chaos is embedded in a broader nonlinear structure involving symmetry breaking, multistability, and chaotic switching \cite{giraldoDrivendissipativeBoseHubbard2020,giraldoChaoticSwitchingDrivendissipative2022}.

These examples establish a useful baseline. In semiclassical open systems, chaos is often organized by identifiable structures in the dissipative phase portrait. This matters for DQC because the corresponding quantum signatures should not be interpreted in isolation. A spectral crossover, a broadened Husimi distribution, a closing Liouvillian gap, or an anomalous trajectory statistic becomes more meaningful when it can be placed near a known bifurcation, attractor merger, or transition between regular and chaotic mean-field motion.

\subsection{Quantum Imprints of Classical Dissipative Structures}\label{quantum-imprints-of-classical-dissipative-structures}

Once the classical phase-space structure is known, the natural question is which quantum observables reflect it. The first family of probes is geometric. Phase-space representations give a direct way to compare a quantum state with the classical attractors, basins, and oscillatory structures of the mean-field flow. In chirally driven dissipative Bose-Hubbard rings, Dahan, Arwas, and Grosfeld showed that a mean-field chaotic attractor is visible in the quantum Wigner function: the quantum weight concentrates in the region of phase space visited by the classical attractor. They also found exponential growth of Lindblad OTOCs, yielding a positive quantum Lyapunov exponent \cite{dahanClassicalQuantumChaos2022}.

Husimi distributions provide a closely related diagnostic in models where coherent states are the natural semiclassical basis. In the open Tavis-Cummings dimer, Mondal, Kolovsky, and Sinha found that chaotic regimes are accompanied by broadened Husimi distributions, rapid decorrelation, suppressed purity, and effective thermal-like behavior of reduced subsystems. Regular regimes, by contrast, retain sharper phase-space structure \cite{mondalDissipativeChaosSteady2025}. This is precisely the type of comparison for which coherent states are useful: they give a phase-space resolution fine enough to reveal whether the quantum state remains localized near regular structures or spreads over a larger chaotic region.

Quantum signatures can also appear in reduced density matrices and in the Liouvillian spectrum. Muraev, Maksimov, and Kolovsky showed that the mean-field transition from a fixed point to a limit cycle in the driven-dissipative Bose-Hubbard dimer produces a visible peculiarity in the stationary single-particle density matrix, and they identified the trimer as the minimal extension where this mechanism can continue into a genuinely chaotic attractor \cite{muraevQuantumManifestationClassical2023}. Dutta, Zhang, and Haque clarified the spectral side of the same problem: fixed-point attractors correspond to a gapped Liouvillian spectrum, limit cycles leave slow-decaying branches with vanishing decoherence rates, and the bifurcation point itself is marked by a spectral collapse \cite{duttaQuantumOriginLimit2025}.

The most systematic steady-state realization of this correspondence so far is the SU(3) Bose-Hubbard trimer studied by Rufo et al. \cite{rufoQuantumSemiClassicalSignatures2025}. Combining exact diagonalization with semiclassical Langevin dynamics, they showed that the sign of the classical Lyapunov exponent organizes the structure of the quantum steady state. When the long-time classical dynamics is regular, with trajectories confined near fixed points or limit cycles, the Husimi distribution remains localized, the entropy stays low, and the effective steady-state Hamiltonian exhibits Poissonian level statistics. When the classical dynamics is chaotic, the steady state delocalizes over phase space, the entropy grows, the Liouvillian gap closes, and the effective steady-state Hamiltonian develops Wigner-Dyson statistics.

This example is important because it connects several diagnostics at once. Strong symmetries constrain the system to lower-dimensional manifolds, suppressing chaos and enforcing localization, while weak symmetries preserve the global phase-space structure and allow chaotic behavior to persist. The phase-space inverse participation ratio defines an effective dimension \(D\) for the support of the Husimi distribution, leading to an entropy scaling of the form \(S\propto \ln N^D\). In this way, geometric delocalization, entropy growth, spectral structure, and classical Lyapunov behavior are tied together in one organized semiclassical picture \cite{rufoQuantumSemiClassicalSignatures2025}.

The broader lesson is that classical dissipative structures can leave clear quantum fingerprints, but those fingerprints are distributed across several observables. Phase-space delocalization, changes in reduced density matrices, entropy growth, Liouvillian gap behavior, and spectral statistics of effective steady-state Hamiltonians all probe different aspects of the same underlying dynamics. A robust diagnosis of semiclassical DQC therefore typically requires combining these observables rather than relying on any single one.

\subsection{Trajectories, Fluctuations, and Transient Chaos}\label{trajectories-fluctuations-and-transient-chaos}

The plurality of diagnostics becomes even more important once one moves from ensemble-averaged density matrices to stochastic quantum trajectories. In open systems, the density matrix can be represented as an average over conditioned realizations, and those realizations can display structures that are partly washed out in the averaged state. This makes trajectory-level chaos a natural, and in some cases indispensable, part of DQC.

Different probes can emphasize different parts of the same semiclassical structure. In the dissipative kicked top, Passarelli et al.~showed that trajectory-averaged nonstabilizerness, or magic, tracks the underlying chaotic mean-field regime more faithfully than entanglement entropy, with the latter losing sensitivity to chaos in the thermodynamic limit \cite{passarelliChaosMagicDissipative2025}. In the Tavis-Cummings dimer, Ray and Kulkarni found that ergodic indicators such as level-spacing ratios and participation measures align with the classical Lyapunov exponent across the phase diagram \cite{rayErgodicChaoticProperties2024}. These results reinforce the idea that semiclassical DQC is not captured by a unique order parameter. Lyapunov exponents, phase-space distributions, entropy growth, reduced density matrices, nonstabilizerness, trajectory statistics, and spectral data each resolve a different projection of the open dynamics.

Ferrari et al.~sharpened this point by arguing that quantum chaos in open systems should be defined operationally at the level of individual conditioned realizations, not only through the asymptotic density matrix. They introduced the spectral-statistics-of-quantum-trajectories (SSQT) criterion to diagnose chaos in trajectory ensembles \cite{ferrariDissipativeQuantumChaos2025}. Applied to driven-dissipative Bose-Hubbard systems, this framework reveals a quantum chaotic phase in parameter regimes where the deterministic mean-field equations and the truncated-Wigner semiclassical limit remain integrable. In those regimes, chaos is generated by fluctuations intrinsic to the quantum jump process rather than inherited from a pre-existing chaotic mean-field attractor.

This is one of the clearest ways in which DQC goes beyond a simple semiclassical correspondence. The semiclassical limit still provides the reference structure, but dissipation and measurement backaction can generate genuinely quantum chaotic behavior that is not already present in the deterministic classical equations. In such cases, coherent-state phase-space methods and mean-field dynamics remain useful, but they must be supplemented by trajectory-level statistics.

A related subtlety is the distinction between transient and asymptotic chaos. Positive local instability at intermediate times does not guarantee that the long-time state is itself chaotic. Mondal, Santos, and Sinha made this explicit in the anisotropic open Dicke model by separating transient chaos, signaled by early-time growth of open-system OTOCs and entropy-like measures, from steady-state chaos diagnosed through long-time observables and effective spectral probes \cite{mondalTransientSteadyStateChaos2026a}. Their results clarify why Liouvillian Ginibre statistics can sometimes diagnose short-time relaxation complexity rather than the character of the final steady state. Richter, S\'a, and Haque reached a compatible conclusion in boundary-driven many-body systems, where the asymptotic structure of the steady state need not mirror the mechanisms governing transient relaxation \cite{richter2025}.

The transient-asymptotic distinction is not merely technical. It determines which diagnostic is appropriate and what dynamical object the diagnostic is actually probing. A Liouvillian spectrum, a steady-state density matrix, a trajectory ensemble, and a finite-time OTOC can all be valid indicators, but they need not answer the same question.

\subsection{From Phase-Space Structure to Quantum Signatures}\label{from-phase-space-structure-to-quantum-signatures}

The semiclassical perspective gives DQC a concrete phase-space language. Fixed points, limit cycles, multistability, bifurcations, and strange attractors provide an organizing picture for open quantum dynamics in large-spin, large-particle-number, or large-occupation limits. Coherent states and phase-space distributions make this picture operational by providing classical-like quantum states whose evolution can be compared directly with dissipative mean-field flows.

At the quantum level, however, the imprint of this structure is distributed across several observables rather than concentrated in a single universal diagnostic. Phase-space distributions, entropy growth, reduced density matrices, Liouvillian spectra, effective steady-state Hamiltonians, trajectory statistics, and fluctuation-sensitive probes can each reveal different aspects of the same underlying dynamics.

The classical-to-quantum correspondence is therefore real but fragile. Classical chaotic attractors can leave clear quantum signatures, but the reverse implication is not guaranteed, and different diagnostics may refer to different time regimes or different dynamical objects.

Dissipation is the main reason for this fragility. It can smear classical structures, suppress chaos, delay its appearance to transient regimes, or generate trajectory-level chaotic behavior beyond what deterministic mean-field equations predict \cite{ferrariDissipativeQuantumChaos2025,dahanClassicalQuantumChaos2022,rufoQuantumSemiClassicalSignatures2025,villasenorCorrespondencePrincipleDissipation2025}.

\section{Waiting-time distributions: a  direct experimental route to DQC?}
\label{Lyapunov}

In the previous section, we argued that individual quantum trajectories open a new window into DQC. In this section, we illustrate this viewpoint using the waiting-time statistics of quantum jumps in quantum trajectories, obtained by unraveling the Lindblad dynamics; see Fig.~\ref{fig:objects}. This idea was proposed in Ref.~\cite{yusipovQuantumLyapunovExponents2019}. It was then tested with a model that describes an open, periodically driven Kerr cavity~\cite{Yusipov2020PhotonWaiting}. In the model, the cavity mode is described by the annihilation operator $a$, the Kerr nonlinearity $\chi$, and a time-periodic coherent drive $F(t)$. The model Hamiltonian is
\begin{equation}
H(t)=\frac{\chi}{2}a^{\dagger 2}a^2+iF(t)(a^\dagger-a),
\label{kerr1}
\end{equation}
while photon losses are described by a single jump operator $L=\sqrt{\gamma}\,a$. The quench-like driving function  was used,
\begin{equation}
F(t)=
\begin{cases}
A, & 0<t\leq T/2,\\[2mm]
0, & T/2<t\leq T,
\end{cases}
\label{kerr2}
\end{equation}
extended periodically as $F(t)=F(t+T)$.
The corresponding Lindblad equation is then unravelled into quantum trajectories. Since there is only one dissipative channel, every quantum jump is identified with the emission of one photon. 

The largest quantum Lyapunov exponent  $\lambda_{\max}^{(\mathrm{QLE})}$, is defined by comparing the evolution of two initially close trajectories~\cite{yusipovQuantumLyapunovExponents2019}. One trajectory is chosen as a fiducial trajectory $\psi_f(t)$, while the second one, $\psi_a(t)$, is initialized as a weakly perturbed auxiliary copy. For a chosen observable $O$, for example, $O=a$ or $O=n=a^\dagger a$ in a cavity, one follows the mismatch $
\Delta(t)=\left|\langle O\rangle_f(t)-\langle O\rangle_a(t)\right|$. Whenever the mismatch reaches a prescribed threshold $\Delta_{\mathrm{max}}$, at time $t_k$, the auxiliary trajectory is brought back close to the fiducial one, so that the distance is reset to a fixed small value $\Delta_0$. We record the times $\mathbf{t}=\{t_1,t_2,...,t_k,\ldots\}$ and estimate the largest quantum Lyapunov exponent defined  as
\begin{equation}
\lambda_{\max}^{\mathrm{QLE}}
=\ln(\frac{\Delta_{\mathrm{max}}}{\Delta_{0}})\cdot
\lim_{t\to\infty}
\frac{K(t)}{t}.
\label{eq:qle_definition}
\end{equation}
In words, it is the rate of clicks weighted with the relative grow factor. A positive value of $\lambda_{\max}^{\mathrm{QLE}}$ signals sensitivity of quantum trajectories to small perturbations and is used to distinguish chaotic from regular dissipative regimes.  For the driven Kerr model, the photon annihilation operator $a$ was used to define the maximal quantum Lyapunov exponent~\cite{Yusipov2020PhotonWaiting}.

\begin{figure}[t]
\centering
\includegraphics[width=0.8\linewidth]{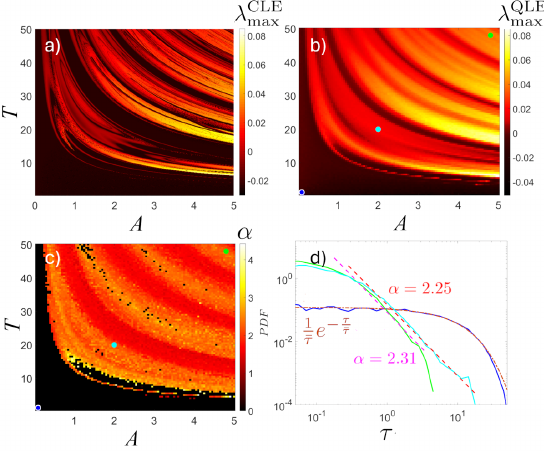}
\caption{
Waiting-time statistics for the driven dissipative Kerr cavity, Eqs.~\eqref{kerr1} and~\eqref{kerr2}~\cite{Yusipov2020PhotonWaiting}.
Largest Lyapunov exponents on the $(A,T)$ parameter plane for the (a) mean-field  and (b) quantum-trajectory dynamics. 
(c) Exponent $\alpha$ of the truncated power-law fit in Eq.~\eqref{eq:truncated_power_law}. Black regions indicate where the fit is rejected. 
(d) Waiting-time distributions for the three marked parameter points, showing nearly Poissonian exponential statistics in the regular regime and intermediate power-law behaviour in the chaotic regime. 
%Adapted from Ref.~\cite{Yusipov2020PhotonWaiting}.
} 
\label{fig:waiting_theory}
\end{figure}

While the reincarnation of quantum Lyapunov exponents in the DQC context is interesting in its own right and serves as yet another bridge to classical dissipative chaos [see Fig.~\ref{fig:waiting_theory}(a,b)], this quantity is not directly accessible in experiments. In contrast, the statistics of waiting times between consecutive photon emissions can be measured directly. The statistics can be defined as follows.

We return to the list $\mathbf{t}$ %$\{t_1,t_2,\ldots\,t_k,\ldots\}$ 
of the detected jump/emission times and use them to define waiting times  $\tau_k=t_{k+1}-t_k$. From the list of waiting times, we define the mean waiting time
$
\bar{\tau}=\langle \tau_k\rangle
$
and the standard deviation
$
\sigma_\tau=
\sqrt{\langle \tau_k^2\rangle-\langle \tau_k\rangle^2}.
$
For a Poisson click process, the waiting-time distribution is exponential,
\begin{equation}
P(\tau)=\frac{1}{\bar{\tau}}
\exp\left(-\frac{\tau}{\bar{\tau}}\right),
\end{equation}
and therefore $\sigma_\tau=\bar{\tau}$. Thus, the ratio $\bar{\tau}/\sigma_\tau$ is equal to one for Poisson statistics. If $\bar{\tau}/\sigma_\tau<1$, then $\sigma_\tau>\bar{\tau}$, and the waiting-time distribution is broader than exponential. This corresponds to what is called in quantum optics 
``super-Poissonian statistics'' \citep{Fox2006QuantumOptics}.

In the chaotic regime of the driven Kerr cavity, the instantaneous photon number $n(t)=\langle a^\dagger a\rangle$ fluctuates strongly along a quantum trajectory. The effective emission rate is therefore not constant. If $\eta_k$ is the random jump threshold, we can write, 
$
\tau_k=-\frac{\ln \eta_k}{s_k}, ~~s_k=\gamma n^{(\mathrm{eff})}_k $.
For regular dynamics, $s_k$ is narrowly distributed and the waiting-time distribution remains close to the exponential Poissonian law. For chaotic dynamics, the broad distribution of $s_k$ produces a non-Poissonian intermediate-time asymptotics. In Ref.~\cite{Yusipov2020PhotonWaiting}, the asymptotics was fitted by a truncated power law,
\begin{equation}
P(\tau)\sim \tau^{-\alpha}\exp(-\tau/\tau_c),
\label{eq:truncated_power_law}
\end{equation}
The parameter regions where this asymptotic was found correlate with the regions where the largest quantum Lyapunov exponent is positive; see Fig.~\ref{fig:waiting_theory}(c).

Recently, Pankratov et al.~\cite{Pankratov2025} recorded switching events induced by thermal microwave photons emitted from a copper 3D resonator using an underdamped Josephson-junction single-photon detector~\cite{Pankratov2025}.  %The detector had a very low dark-count rate and was operated in the few-photon regime. 
%The relevant resonator modes were near $8.8\,\mathrm{GHz}$ and $14.0\,\mathrm{GHz}$, with higher modes near $16.7\,\mathrm{GHz}$ and $17.9\,\mathrm{GHz}$ contributing much less below $80\,\mathrm{mK}$; see Fig.~\ref{fig:modes_temperature}. 
The resonator temperature was varied from the base-temperature regime to about $80\,\mathrm{mK}$, and the click statistics was measured as a function of temperature. The corresponding waiting-time distributions were reconstructed from the intervals between consecutive click events.

At the lowest temperatures, thermal photons are very rare. The observed clicks are then dominated by the detector dark counts, which are essentially random escape events of the Josephson detector. Their statistics is therefore close to Poissonian, and $\bar{\tau}/\sigma_\tau\approx1$. When the temperature is increased, the first resonator mode becomes thermally populated and dominates the detected photon flux. A single thermal bosonic mode has Bose--Einstein photon-number statistics, with $(\Delta n)^2=\bar n+\bar n^2$.
The additional term $\bar n^2$ gives photon bunching. In waiting-time language, the distribution becomes broader than exponential, so that $\sigma_\tau>\bar{\tau}$ and $\bar{\tau}/\sigma_\tau<1$. This is the origin of the observed super-Poissonian regime~\cite{Pankratov2025}. 
At  higher temperatures, the second resonator mode starts to contribute substantially. The detector then sees radiation from more than one thermal mode. For $N_m$ independent thermal modes, the excess fluctuation is reduced, 
$(\Delta n)^2=\bar n+\frac{\bar n^2}{N_m}$.
Thus increasing the number of contributing modes moves the statistics closer to the Poissonian limit.

\begin{figure}[t]
\centering
\includegraphics[width=0.99\linewidth]{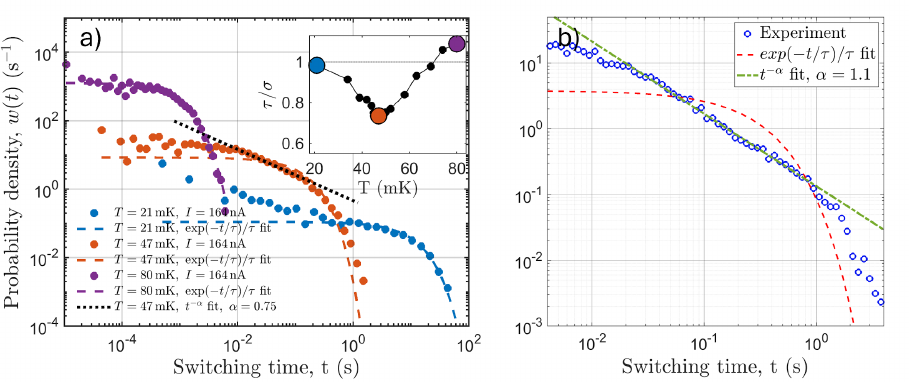}
\caption{\textbf{Waiting-time statistics observed in experiments with a microwave cavity~\cite{Pankratov2025}}. (a) Distribution of times between sequential switchings of the single-photon detector at various temperatures. (b) The same as (a), but with a different coupling between the antenna and the cavity. Experimental data courtesy of A. Pankratov.}
\label{fig:modes_temperature}
\end{figure}

Although Ref.~\cite{Yusipov2020PhotonWaiting} was referenced and mentioned throughout Ref.~\cite{Pankratov2025}, and although the idea of using the intermediate power-law fit was used to analyse the experimental data, the connection between the observed statistics and DQC should be stated carefully.

In the theoretical Kerr-cavity model~\cite{Yusipov2020PhotonWaiting}, the truncated power-law waiting-time distribution appears in parameter regions corresponding to mean-field chaotic regimes and correlates with positive classical and quantum Lyapunov exponents. There, the mechanism is genuinely dynamical, since chaotic quantum trajectories generate strongly fluctuating effective emission rates. In the microwave experiment~\cite{Pankratov2025}, the immediate explanation is more standard quantum optics. The single-mode thermal light is super-Poissonian, while the multimode thermal light is closer to Poissonian. Therefore, the experiment is not yet a direct demonstration of  DQC features similar to the ones observed with the cavity model.  A direct DQC claim would require the construction of an effective model of the coupled resonator--Josephson-detector system, which is itself nonlinear, together with chaos diagnostics for this model, using, e.g., spectral statistics.

A further observation made in experiments~\cite{PankratovPrivateCommunication2026} brings us even closer to the DQC setting. At a fixed low temperature of the resonator, where thermal photons are very rare, an external microwave signal, with tunable parameters,  was applied. In this driven regime, the detector click statistics changes from Poissonian to strongly super-Poissonian, with $\bar{\tau}/\sigma_\tau$ dropping to $0.5$, compared with the minimum value of about $0.63$ observed for thermal photons. Again, this externally controlled crossover is not yet described by a model (which is still missing), but it is suggestive because it combines a coherent driving, cavity, and a nonlinear Josephson detector in the same setting. It therefore provides a  motivation for developing an effective open-system model of the driven resonator--detector circuit and testing it with the spectral and trajectory-based DQC diagnostics.

\section{Grobe-Haake-Sommers model: 38 years later}
\label{grobe_revision}

We now return to the starting point of DQC and examine more carefully the Grobe--Haake--Sommers model~\cite{grobeQuantumDistinctionRegular1988}. Our novel observation is that, in the regular case $k_1=0$, the model has an additional fine structure because its operator space, and therefore its spectrum, decomposes into smaller invariant sectors, which Ref.~\cite{grobeQuantumDistinctionRegular1988} missed.

\begin{figure}[t]
    \centering
    \includegraphics[width=0.95\textwidth]%{figures/Figure12.pdf}
    {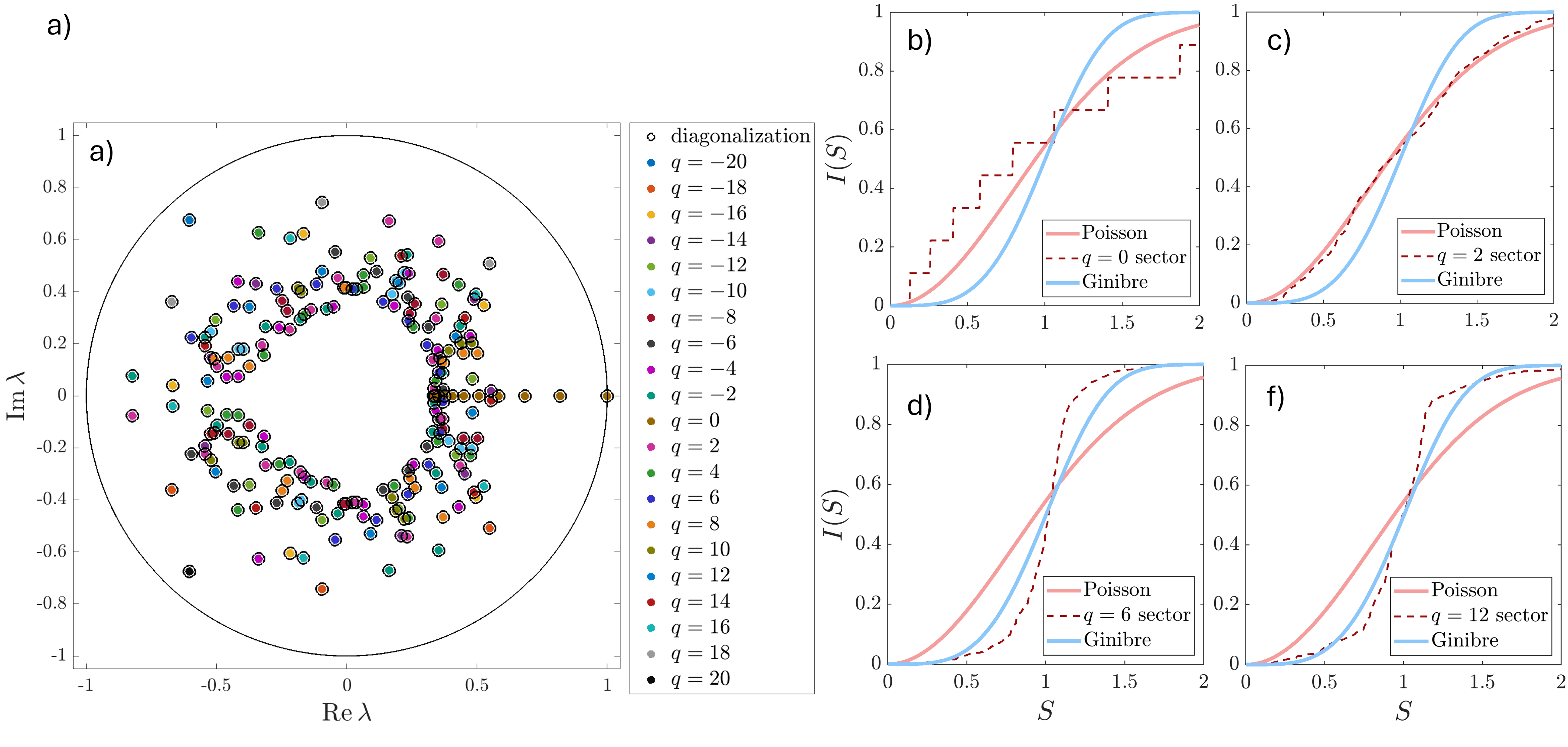}
    \caption{Spectral properties of the  Grobe--Haake--Sommers map in the regular regime~\cite{grobeQuantumDistinctionRegular1988}, $k_1=0$. The parameters are $S=10$, $\Gamma=0.1$, $p=2$, and $k_0\in[10,12]$, as in the original work. Panel (a) shows the eigenvalues in the even parity block for $k_0=10$. Open black circles are obtained by direct diagonalization of the map, while coloured points are obtained from the explicit fixed-$q$ formula~\eqref{eq:grobe_explicit_eigenvalues_q} (different colours correspond to different $q$ sectors). The two spectra coincide up to precision $\sim 10^{-15}$. Panels (b)--(e) show integrated nearest-neighbour spacing distributions for different fixed-$q$ sectors, compared with the Poisson and Ginibre reference curves.
    }
    \label{fig:grobe_q_sectors}
\end{figure}

In the basis $|m\rangle\langle n|$, the Hamiltonian and dissipative parts preserve the difference $q=m-n$. In modern language, we say that $J_z$ generates a weak symmetry~\cite{buca2012} of the Lindbladian $\cL_{H_0}+\cL_D$. Hence, each parity block splits further into invariant fixed-$q$ sectors,
$K_q=\mathrm{span}\{|m\rangle\langle n|:\ m-n=q\}$.
Writing $n=m-q$, the allowed values of $m$ are
$m=\max(-S,-S+q),\ldots,\min(S,S+q)$, and the dimension of the sector is $d_q=2S+1-|q|$.
Different fixed-$q$ sectors thus contain different numbers of eigenvalues. For example, for $S=10$, one has $d_0=21$, $d_2=19$, $d_6=15$, and $d_{12}=9$.
The fixed-$q$ blocks are triangular, so their eigenvalues are given directly by the diagonal entries. For $n=m-q$, the eigenvalues can be written explicitly,
\begin{equation}
\label{eq:grobe_explicit_eigenvalues_q}
\lambda_m^{(q)}(k_0)
=
\exp\!\left[
-\frac{\Gamma}{2S}
\left(a_m+a_{m-q}\right)
\right]
\exp\!\left[
-iq\left(
p+\frac{k_0}{2S}(2m-q)
\right)
\right],
\end{equation}
where $
a_m
=
S(S+1)-m(m-1)
=
(S+m)(S-m+1)
$. Equivalently,
\begin{equation}
\lambda_m^{(q)}(k_0)=r_m^{(q)}\exp\left[i\theta_m^{(q)}(k_0)\right],
\end{equation}
with
\begin{equation}
r_m^{(q)}
=
\exp\!\left[
-\frac{\Gamma}{2S}
\left(a_m+a_{m-q}\right)
\right],~~ \theta_m^{(q)}(k_0)
=
-q\left(
p+\frac{k_0}{2S}(2m-q)
\right).
\end{equation}
The radius depends on $m$ and $q$, but not on $k_0$. Varying $k_0$ changes only the angular positions of the eigenvalues.
The $q=0$ sector is exceptional and has all eigenvalues real, positive, and independent of $k_0$,
\begin{equation}
\label{eq:grobe_q0_eigenvalues}
\lambda_m^{(0)}
=
\exp\!\left[-\frac{\Gamma}{S}a_m\right].
\end{equation}

Figure~\ref{fig:grobe_q_sectors}(a) gives a direct check of this analytic structure. The eigenvalues obtained by diagonalizing the full even parity block coincide (up to precision $10^{-15}$) with the union of eigenvalues obtained from Eq.~\eqref{eq:grobe_explicit_eigenvalues_q} over the corresponding fixed-$q$ sectors.
Thus, for $k_1=0$, the Poisson statistics of the full parity block appears as a result of combining many independent fixed-$q$ sectors into one sector and ignores the block decomposition.
For a fixed $k_0$, the nearest-neighbour spacing statistics should be computed separately for each value of $q$. One should first construct finite set $\{\lambda_m^{(q)}(k_0)\}_m$, unfold this spectrum, compute the nearest-neighbour spacings inside this single spectrum, and only then add the result to the accumulated histogram.

The difference between the fixed-$q$ sectors is already visible at the level of pair distances. For two labels $m$ and $m'$, one has
\begin{equation}
\left|
\lambda_m^{(q)}
-
\lambda_{m'}^{(q)}
\right|^2
=
\left(r_m^{(q)}\right)^2
+
\left(r_{m'}^{(q)}\right)^2
-
2r_m^{(q)}r_{m'}^{(q)}
\cos\Delta\theta_{m,m'}^{(q)},~~\mathrm{where}~~ \Delta\theta_{m,m'}^{(q)}
=
-\frac{k_0q}{S}(m-m').
\end{equation}
A small spacing therefore requires two conditions at the same time, (i) the two radii must be close and (ii) the angular phase difference must be close to a multiple of $2\pi$. Changing $q$ changes both the radial profile $r_m^{(q)}$ and the angular arithmetic in $\Delta\theta_{m,m'}^{(q)}$. This explains why the integrated spacing distributions in Fig.~\ref{fig:grobe_q_sectors}(b)--(e) can look different for different sectors, even though all of them are generated by the same explicit integrable formula. The sector-resolved spectra thus reveal the finer deterministic structure hidden in the parity-block partition.

This point is important for the interpretation of apparent `level repulsion' in fixed-$q$ sectors. When, for example, a sector such as $q=6$ or $q=12$ produces an integrated spacing distribution closer to the Ginibre curve than to the Poisson curve, this should not be interpreted as a genuine random-matrix mechanism. The eigenvalues are still deterministic functions of $m$ and $k_0$. The apparent suppression of small spacings is a geometric effect caused by the very specific mechanism of $q$-dependent radii and angular velocities. In contrast, the $q=0$ sector is a purely real sequence and should not be compared to a two-dimensional point process as a matter of principle.

We conclude that the $k_1=0$ point of the Grobe-Haake-Sommers model is, in a sense, singular, with all eigenvalues related to the parameters of the model without the need for diagonalization (which lies at the origin of random-matrix behaviour). One should, instead, do the level analysis of the ``regular regime'' at a small nonzero value of $k_1$, where the weak $J_z$ symmetry is broken. The main lesson to take from this revision is that a careful symmetry analysis and resolving all symmetries is crucial for a proper application of the DQC conjecture.

\section{`Challenges' to the DQC surmise: potential and current}
\label{outlook}

DQC is not a mathematical theory, and it is naive to expect it to provide universally valid definitions of chaotic and regular regimes of open quantum dynamics. Whatever diagnostic, measure, or quantifier one adopts, counterexamples can be constructed. In this respect, the situation is no different from conventional Hamiltonian quantum chaos, where the distinction between integrable and chaotic quantum systems is also operational rather than absolute, and where the very notion of quantum integrability requires care~\citep{Caux2011}. Nevertheless, this does not make the DQC concept empty. The value of DQC lies in identifying robust and reproducible \textit{signatures}, which help to consistently distinguish large classes of regular and chaotic dissipative systems.

The kicked-top analysis of Grobe, Haake, and Sommers established the original semiclassical correspondence principle for DQC, according to which classical chaotic attractors should imprint Ginibre-like statistics on the Lindbladian spectrum. However, it is now clear that the correspondence is not universal. In the open Dicke model and the dissipative kicked top, Villase\~nor and collaborators showed that Ginibre-type correlations can appear even when the classical dynamics has only simple attractors, and that chaotic classical motion does not always generate the expected spectral signature~\cite{villasenorBreakdownQuantumDistinction2024,villasenorCorrespondencePrincipleDissipation2025,villasenorAnalysisChaosRegularity2024a}. The lesson is not that the spectral DQC viewpoint fails, but that no single spectral observable provides a universal one-to-one proxy for the underlying classical dynamical regime.

A related source of apparent controversy concerns the relation between spectral statistics of a Lindbladian and spectral statistics of the corresponding steady state. The spectrum of the Lindbladian, $\{\ell_\alpha\}$, controls the decay modes so that the state of the evolving system at  time $t$ is given by
\begin{equation}
\rho(t)=\rho_{\mathrm{ss}}+\sum_{\mathrm{Re}~ \ell_\alpha < 0}c_\alpha e^{\ell_\alpha t} r_\alpha,
\label{eq:modes}
\end{equation}
whereas the spectrum of the steady state $\rho_{\mathrm{ss}}$, or of an effective steady-state Hamiltonian constructed from it, probes only the final invariant state. Therefore, a Lindbladian spectrum with Ginibre-like CSR, including an annular or ``bitten-donut'' structure, can coexist with a steady state whose eigenvalue statistics are close to Poissonian. Conversely, a steady state may show random-matrix-like signatures even when the spectrum of decay modes is not Ginibre-like. This should not be regarded as an internal contradiction of DQC. It says that transient relaxation and the final attractor are different facets of open-system dynamics.

The classical analogy is direct. In dissipative classical systems, chaotic transient motion may end on a simple fixed point or a limit cycle.
Conversely, a trajectory may show regular-looking finite-time motion before it reaches a strange attractor. There may also be regular relaxation to a regular attractor and chaotic relaxation to a strange attractor. A particular scenario can change sharply under a crisis, e.g., when a chaotic attractor or its basin collides with an unstable invariant set or a basin boundary, producing a sudden change of the attractor while leaving parts of the finite-time dynamics almost unchanged~\cite{grebogi1983crises}. The quantum problem is to identify which of these classical scenarios is represented by a given combination of CSR and steady-state statistics, trajectory diagnostics, and phase-space structure. The mismatch between the spacing ratios of the Lindbladian and the asymptotic state it pulls to is therefore not a failure of the spectral DQC surmise, but a sign that different metrics resolve different time sectors of open quantum evolution.

The same applies to different sectors of operator space. One may project the GKLS dynamics of a specific model onto a selected subset of Lindbladian modes, see Eq.~\eqref{eq:modes}, and then unravel the corresponding contribution into an ensemble of trajectories ~\cite{ferrariDissipativeQuantumChaos2025,richter2025localization}. Obviously, this ensemble need not display the same degree of chaoticity (or regularity) as signaled by statistical features of the full Lindbladian spectrum. This should not be viewed as an inconsistency. It is the quantum counterpart of a familiar feature of classical dissipative chaos, which is generally not distributed uniformly over phase space~\cite{Ott2002}. Except in idealized uniformly hyperbolic systems, such as Anosov flows~\cite{Ott2002}, different regions of phase space can have different local Lyapunov exponents, include stable and unstable manifolds, transient chaotic sets, and basins of attraction. Thus, by selecting different spectral modes and the corresponding trajectory ensembles, one can observe different local manifestations of chaoticity without invalidating the much broader spectral DQC picture.

A second kind of apparent ``violation'' of the DQC spectral conjecture concerns the comparison between, on the one hand, the deterministic classical equations obtained in the mean-field limit or lowest-order semiclassical stochastic extensions, and, on the other hand, the genuinely quantum  dynamics. In the driven-dissipative Bose--Hubbard dimer, the mismatches reported in Ref.~\cite{ferrariDissipativeQuantumChaos2025} occur mainly in the direction in which classically regular motion, such as relaxation to a fixed point or limit cycle, is replaced by chaotic trajectory dynamics in the quantum limit.

We do not see this as a contradiction to the DQC picture. Rather, such observations indicate that the relevant finite-$N$ quantum dynamics is not fully represented by low-order stochastic semiclassical approximations. Environment-induced jumps and finite-particle-number fluctuations  prevent trajectories from settling onto the corresponding set of classical  trajectories -- even in the statistical sense -- and can therefore generate chaotic trajectory-level dynamics in parameter regions where the classical limit is regular. The fact that these discrepancies are concentrated near boundaries between the classical regular and chaotic regions~\cite{ferrariDissipativeQuantumChaos2025} further supports our interpretation. These are precisely the regions where the dynamics is expected to be most sensitive to finite-size fluctuations and/or higher-order semiclassical corrections.

In this sense, the issue is not  whether the classical mean-field or truncated Wigner approximation (TWA) models are chaotic (or regular) for the same set of parameters as the original quantum model. The question is whether the semiclassical stochastic process correctly reflects the finite-$N$ quantum dynamics. The TWA does introduce noise, but only after reducing the exact phase-space evolution to a lowest-order Fokker--Planck description and discarding higher-order terms. This approximation therefore misses parts of the complex fluctuation structure, which may well be responsible for the observed quantum chaoticity. This was explicitly noted by Ferrari, Minganti, Aron, and Savona in their later work~\cite{ferrariChaoticQuantumDynamics2025}. It is also illustrated by the recent semiclassical analysis of Rufo, Rufo, Ribeiro, and Chesi~\cite{rufoQuantumSemiClassicalSignatures2025}, where quantum fluctuations do not enter merely as an additive Gaussian noise but, on the level of the next-to-leading corrections, result in a multiplicative stochastic term that depends explicitly on the phase-space coordinates. Thus, the reported ``violation''~\cite{ferrariDissipativeQuantumChaos2025} can be viewed as yet another observation of boundary-region effects in the quantum-to-classical correspondence, where the  hierarchy of fluctuation corrections becomes relevant~\cite{habib2004,greenbaum2005,chou2011,ivanchenko2017}.

Finally, Ref.~\cite{navesLevelRepulsionFails2026} raises a different issue. The authors consider integrable open models, which, after a finite Hilbert-space truncation or after imposing open boundaries, may display Ginibre-like level repulsion once the dimension of the model Hilbert space exceeds a certain threshold. They attribute this effect to strong non-normality, whereby bulk eigenvalues become highly ill-conditioned and the corresponding eigenvectors develop a non-Hermitian skin effect in Liouville space.
This is a valid warning, but its implication should be formulated more narrowly. In our opinion, the key message of the paper can be distilled into the following statement: When diagonalizing non-normal operators or matrices numerically, one should do it with care, since the eigenproblem may be strongly ill-conditioned. This observation certainly has didactic value (and was also presented and discussed in recent works, see, for example, Refs.~\cite{Znidari2022Solvable,Znidari2024Momentum}),
but it can hardly be regarded as a challenge to the DQC spectral conjecture.
We checked some spin models considered in Ref.~\cite{saComplexSpacingRatios2020a} and found that, in the regular regimes, the eigenproblems are, in fact, well conditioned, namely, the relevant condition numbers grow only polynomially with the dimension of the Hilbert space. 
%Strong non-normality might appear in the DQC context and indeed might be a source of possible numerical or boundary-induced artifacts, but it does not, by itself, constitute a general objection to the DQC surmise.

%%%%%%%%%%%%%%%%%%%%%%%%%%%%%%%%%%%%%%%%%%%%%%%%%%%%%%%%%%%%%

\bibliographystyle{JHEP}
\bibliography{Pedro_bib_file}

\end{document}